\documentclass[twocolumn,epjc3]{svjour3}

\usepackage[utf8]{inputenc} 
\usepackage[T1]{fontenc} 	
\usepackage[main=english, french]{babel} 

\usepackage[labelfont=bf]{caption}
\usepackage{amsmath} 		
\usepackage{amsfonts}
\usepackage{mathtools} 		
\usepackage{float} 			
\usepackage{graphicx} 		
\usepackage{tabularx} 		
\usepackage{booktabs} 		
\usepackage{color, xcolor} 	
\usepackage{pdfpages} 		
\usepackage{extarrows} 		
\usepackage{multirow} 		
\usepackage{enumitem} 		
\usepackage{xspace} 		
\usepackage{stackrel} 		
\usepackage{tikz} 			
\usepackage{braket} 		
\usepackage{bm} 			
\usepackage{tensor} 		
\usepackage{slashed} 		
\usepackage{siunitx} 		
\usepackage{cite} 			
\usepackage[normalem]{ulem} 
\usepackage{titlesec} 		
\usepackage{doi} 			
\usepackage{hyperref} 		
\usepackage[nottoc, notlot, notlof]{tocbibind} 	
\usepackage[notmath]{sansmathfonts} 

\usepackage{algorithmic}
\usepackage[ruled, vlined]{algorithm2e} 
\SetKwInOut{Input}{Input}
\SetKwInOut{Output}{Output}
\SetKwFunction{Fforward}{Forward}
\SetKwFunction{Fbackward}{Backward}
\SetKwProg{Fn}{function}{:}{}
\DontPrintSemicolon

\definecolor{EmeraldGreen}{HTML}{1ea78d}
\definecolor{EnglishRed}{HTML}{b02427}
\hypersetup{
    pdftitle={Elsa},
	pdfauthor={Nachman et al.},
    colorlinks=true,
    urlcolor=EmeraldGreen,
    citecolor=EmeraldGreen,
    linkcolor=EnglishRed}

\setitemize{itemsep=2pt,topsep=2pt,parsep=0pt,partopsep=0pt,leftmargin=*}
\setenumerate{itemsep=2pt,topsep=2pt,parsep=0pt,partopsep=0pt,leftmargin=*}
\newcommand{\mailben}{\href{mailto:bpnachman@lbl.gov}{bpnachman@lbl.gov}}
\newcommand{\mailramon}{\href{mailto:ramon.winterhalder@uclouvain.be}{ramon.winterhalder@uclouvain.be}}

\usepackage{pifont}
\newcommand{\cmark}{\ding{51}}%
\newcommand{\xmark}{\ding{55}}%

\newcommand{\ie}{\text{i.e.}\;}

\newcommand{\mwith}{\text{with}}
\newcommand{\mand}{\text{and}}

\def\d{\mathrm{d}}

\newcommand{\atanh}{\operatorname{atanh}}

\newcommand{\one}{\leavevmode\hbox{\small1\normalsize\kern-.33em1}}

\newcommand{\madgraph}{\textsc{MadGraph5\_aMC@NLO}\xspace}

\newcommand{\pytorch}{\textsc{PyTorch}\xspace}

\newcommand{\lhapdf}{\textsc{LHAPDF6}\xspace}

\newcommand{\augflow}{\textsc{AugFlow}\xspace}
\newcommand{\rambo}{\textsc{Rambo}\xspace}
\newcommand{\rambodiet}{\textsc{RamboOnDiet}\xspace}
\newcommand{\mahambo}{\textsc{Mahambo}\xspace}
\newcommand{\laser}{\textsc{Laser}\xspace}
\newcommand{\elsa}{\textsc{Elsa}\xspace}
\newcommand{\survae}{\textsc{SurVAE}\xspace}
\newcommand{\dctr}{\textsc{Dctr}\xspace}
\newcommand{\dctrgan}{\textsc{DctrGAN}\xspace}
\newcommand{\omnifold}{\textsc{OmniFold}\xspace}

\newcommand{\elfs}{\textsc{Elfs}\xspace}
\newcommand{\prep}{\textsc{Precisesiast}\xspace}

\newcommand{\arXiv}[2][]{%
	\ifthenelse{\equal{#1}{}}%
	{\href{http://arxiv.org/abs/#2}{arXiv:#2}}%
	{\href{http://arxiv.org/abs/#2}{arXiv:#2~[#1]}}}

\def\slashchar#1{\setbox0=\hbox{$#1$}           
   \dimen0=\wd0                                 
   \setbox1=\hbox{/} \dimen1=\wd1               
   \ifdim\dimen0>\dimen1                        
      \rlap{\hbox to \dimen0{\hfil/\hfil}}      
      #1                                        
   \else                                        
      \rlap{\hbox to \dimen1{\hfil$#1$\hfil}}   
      /                                         
   \fi}

\def\mathswitchr#1{\relax\ifmmode{\mathrm{#1}}\else$\mathrm{#1}$\xspace\fi}
\def\mathswitch#1{\relax\ifmmode#1\else$#1$\xspace\fi}

\newcommand{\PW}{\mathswitchr W}
\newcommand{\PWp}{\mathswitchr {W^+}}

\newcommand{\Pp}{\mathswitchr p}

\newcommand{\Pj}{\mathswitchr j}

\newcommand{\jets}{\mathrm{jets}}

\numberwithin{equation}{section} 

\usepackage[shortcuts]{extdash}
\begin{document}
\title{\elsa\ -- Enhanced latent spaces for improved collider simulations}

\preprint{IRMP-CP3-23-20} 
\journalname{Eur. Phys. J. C}

\author{
    Benjamin Nachman\thanksref{e1,inst1, inst2} and  
    Ramon Winterhalder\thanksref{e2,inst3}}
\thankstext{e1}{E-mail: \mailben}
\thankstext{e2}{E-mail: \mailramon}

\institute{Physics Division, Lawrence Berkeley National Laboratory, Berkeley, CA 94720, USA\label{inst1} \and 
Berkeley Institute for Data Science, University of California, Berkeley, CA 94720, USA\label{inst2} \and
CP3, Universit\'e Catholique de Louvain, B-1348 Louvain-la-Neuve, Belgium\label{inst3}}
\date{Received: date / Revised version: date}
%

\maketitle

\begin{abstract}
    Simulations play a key role for inference in collider physics. We explore various approaches for enhancing the precision of simulations using machine learning, including interventions at the end of the simulation chain (reweighting), at the beginning of the simulation chain (pre-processing), and connections between the end and beginning (latent space refinement). To clearly illustrate our approaches, we use W+jets matrix element surrogate simulations based on normalizing flows as a prototypical example. First, weights in the data space are derived using machine learning classifiers. Then, we pull back the data-space weights to the latent space to produce unweighted examples and employ the Latent Space Refinement (\laser) protocol using Hamiltonian Monte Carlo. An alternative approach is an augmented normalizing flow, which allows for different dimensions in the latent and target spaces. These methods are studied for various pre-processing strategies, including a new and general method for massive particles at hadron colliders that is a tweak on the widely-used \rambodiet mapping. We find that modified simulations can achieve sub-percent precision across a wide range of phase space.
\end{abstract}

\tableofcontents

\section{Introduction}
\label{sec:intro}

A cornerstone of modern high-energy collider physics is first-principles simulations that encode the underlying physical laws. These simulations are crucial for performing and interpreting analyses, linking theoretical predictions and experimental measurements. However, the most precise simulations are computationally demanding and may become prohibitively expensive for data analysis in the era of the High-Luminosity Large Hadron Collider (HL-LHC)~\cite{CERN-LHCC-2017-020,CERN-LHCC-2017-014}. As a result, various fast simulation approaches have been developed to mimic precise simulations with only a fraction of the computational overhead.

In the language of machine learning (ML), fast simulations are called \textit{surrogate models}. While most fast simulations use some flavor of machine learning, a growing number of proposed surrogate models are based on deep learning.  Deep generative models include Generative Adversarial Networks (GAN)~\cite{Goodfellow:2014:GAN:2969033.2969125,Creswell2018}, Variational Autoencoders~\cite{kingma2014autoencoding,Kingma2019}, Normalizing Flows~\cite{10.5555/3045118.3045281,Kobyzev2020}, and Diffusion Models~\cite{diff1, diff2, diffvsgan} which often make use of score matching~\cite{scorematching05,scorematching19,scorematching20}.

In collider physics, deep learning-based surrogate models have been proposed for all steps in the simulation chain, including phase-space sampling, amplitude evaluation, hard-scatter processes, minimum bias generation, parton shower Monte Carlo, and detector effects (see e.g. Ref.~\cite{Butter:2022rso,Adelmann:2022ozp} for recent reviews).  In this paper, we focus on the emulation of hard-scatter event generation~\cite{Butter:2019cae,Alanazi:2020klf,Alanazi:2020jod,Bravo-Prieto:2021ehz,Butter:2021csz}, although the approaches are more general.

Simulations can be viewed as deterministic functions from a set of fixed and random numbers into structured data. The fixed numbers represent physics parameters like masses and couplings, while the random numbers give rise to outgoing particle properties sampled from various phase space distributions. For deep generative models, the random numbers are typically sampled from simple probability densities like the Gaussian or uniform probability distributions. The random numbers going into a simulation constitute the \textit{latent space} of the generator. A key challenge with deep generative models is achieving precision. One way to improve the precision is to modify the map from random numbers to structure. A complementary approach is modifying the probability density of the generated data space or the latent space. This approach is what we investigate in this paper.

For a variety of applications, normalizing flows have shown impressive precision for HEP applications~\cite{Verheyen:2020bjw,Krause:2021ilc,Butter:2021csz,Krause:2021wez,Winterhalder:2021ngy,Verheyen:2022tov,Heimel:2022wyj}.However, these models suffer from topological obstructions due to their bijective nature~\cite{dupont2019augmented, wu2020stochastic, Winterhalder:2021ave} which for instance becomes relevant for multi\-/resonant processes and at sharp phase-space boundaries that are often introduced by kinematic cuts. Topological obstructions are not just affecting normalizing flows but can also limit the performance of autoencoders~\cite{Batson:2021agz}. If the phase-space structure is known, the problematic phase-space regions can be either smoothed by a local transformation~\cite{Butter:2021csz} or the flow can be directly defined on the correct manifold~\cite{Brehmer:2020vwc}. However, we generally do not know where these problematic regions lie in the highly complex and multi-dimensional phase space. Moreover, even if we identify and fix problematic patterns in a lower-dimensional subspace, we are not guaranteed to cure all higher-dimensional topological structures. Therefore, we study two complementary strategies to improve the precision of generative networks affected by topological obstructions and other effects. These strategies are both generalizable and computationally inexpensive.

As a first method, we use the \laser protocol~\cite{Winterhalder:2021ave} which relies on the refinement of the latent space of the generative model. The \laser protocol is an extension of the \dctrgan~\cite{2009.03796} method that is not specific to GANs and circumvents the limited statistical power of weighted events by propagating the weights into the latent space. While this idea has also been proposed solely for GANs in Ref.~\cite{che2020gan}, the \laser approach works for any generative model and further extends Ref.~\cite{che2020gan} by allowing more sophisticated sampling techniques.

In the second method, we employ augmented normalizing flows~\cite{dupont2019augmented,huang2020augmented,chen2020vflow,nielsen2020survae}. In contrast to standard normalizing flows, the feature space is augmented with additional variables such that the latent space dimension can be chosen to have higher dimensions than the original feature space dimension. As a result, possibly disconnected patches in the feature space, which are difficult to describe by deep generative models, can be connected by augmenting additional dimensions and thus simplify the mapping the network has to learn. In order to obtain the original feature space, we use a lower\-/dimensional slice of the generated feature space.

\begin{figure}[b]
    \centering
    \includegraphics[width=0.23\textwidth]{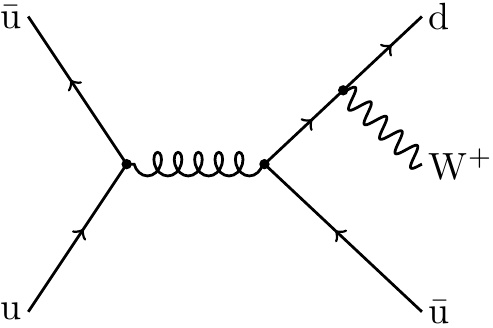}
    \includegraphics[width=0.23\textwidth]{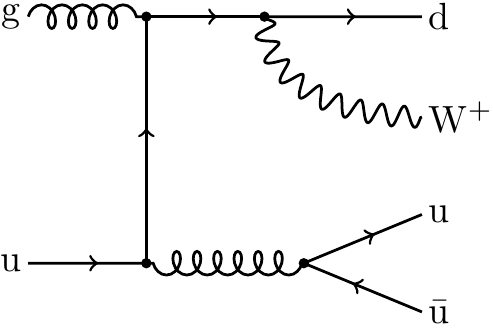}
    \caption{Examples of LO Feynman diagrams contributing to the $\Pp \Pp \to \PWp + \jets$ process.}
    \label{fig:feynman}
\end{figure}

The \laser protocol and the augmented flow show excellent results when applied to simple toy examples. In this paper, we apply both methods to a complicated high-energy physics example for the first time and show how they alleviate current limitations in precision while being conservative regarding network sizes and the number of parameters. Note that augmentation only describes one possible incarnation of a surjective transformation, of which some have already applied on HEP data successfully in Ref.~\cite{Verheyen:2022tov}. In our studies, we consider W-boson production with associated jets~\cite{CMS:2011jak,ATLAS:2012ikx,Hoche:2012tae,Boughezal:2016dtm,LHCb:2016nhs,CMS:2017gbl,ATLAS:2017irc,Gehrmann-DeRidder:2018kng} at the LHC, \ie
\begin{align}
    \Pp \Pp \to \PWp + \jets,
\end{align}
as illustrated in Fig.~\ref{fig:feynman}. Owing to the high multiplicity of the multi-jet process and phase space restrictions (e.g.\ jet $p_{\mathrm{T}}$ thresholds), the associated phase space is large and topologically complex, with many non-trivial features that need to be correctly addressed by the generative models.

This paper is organized as follows. In Sec.~\ref{sec:methods}, we introduce and compare both methods and describe their implementation. We define and list different data representations in Sec.~\ref{sec:data_representation} and introduce the \mahambo algorithm. As a first application, we show simple 2\-/dimensional toy examples to illustrate our approaches in Sec.~\ref{sec:experiments}. Afterwards, we employ both methods on a fully comprehensive LHC example and investigate the impact of various data representations on the models' performance. Finally, we conclude in Sec.~\ref{sec:conclusion}.

\section{Methods}
\label{sec:methods}

The baseline method for both of our models is a normalizing flows\cite{10.5555/3045118.3045281,Kobyzev2020}. Normalizing flows rely on the change of variables formula
\begin{equation}
    \log p_{g(\mathcal{Z})}(x)=\log p_\mathcal{Z}(g^{-1}(x)) + \log \left\vert\frac{\partial g^{-1}(x)}{\partial x}\right\vert\,,
    \label{eq:flow_trans}
\end{equation}
which explicitly encodes an estimate of the probability density $p_{g(\mathcal{Z})}\approx p_\mathcal{X}$ by introducing a bijective mapping $g$ (parameterized as a neural network) from a latent space $\mathcal{Z}=\{z\in\mathbb{R}^N\vert z\sim p_\mathcal{Z}\}$ to a feature space with data described by $\mathcal{X}=\{x\in\mathbb{R}^N\vert x\sim p_\mathcal{X}\}$. In contrast, for VAEs and GANs, the mapping $g$ does not need to be bijective. However, VAEs only yield a lower bound of the density (evidence lower bound, or ELBO) and for GANs the posterior is only defined implicitly.  

\begin{figure}[h!]
    \centering
    \includegraphics[width=0.47\textwidth]{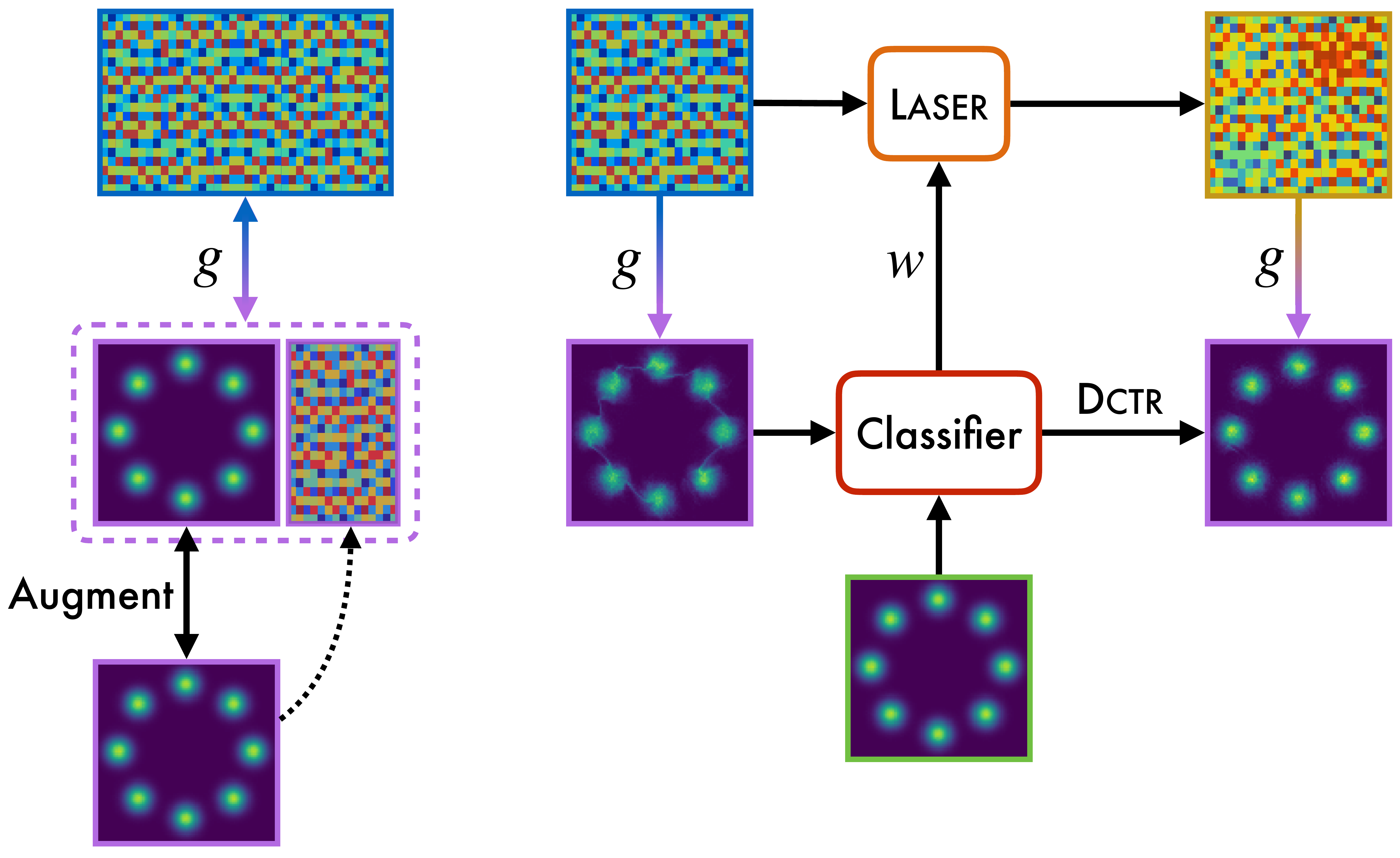}
    \caption{A schematic diagram illustrating the augmented flow (left) and the \laser protocol (right).}
    \label{fig:schematic}
\end{figure}

We study two methods that aim to improve the precision and expressiveness of the generative model $g$ by either modifying the latent space prior $p_\mathcal{Z}$ or by post-processing the data space, either with importance weights or by augmenting the feature space $\mathcal{X}$ with an additional variable $r\sim q(r\vert x)$.
These methods are illustrated in Fig.~\ref{fig:schematic} and are described in more detail in the following.

\subsection{Latent space refinement}
\label{sec:laser}

As the first method, we use the \laser protocol~\cite{Winterhalder:2021ave}, illustrated on the right of Fig.~\ref{fig:schematic}. The \laser method acts as a post-hoc procedure modifying the latent space prior $p_\mathcal{Z}$ to improve the generated samples $x$ while keeping the generative model $g$ fixed. For this method, the baseline generator $g(z)$ can be any generative model or physics simulator (as long as we have access to the input random numbers) and does not have to be bijective in general. Throughout this paper, we choose $g$ as a normalizing flow for simplicity and to make this method better comparable with the second method.

As a first step in the \laser protocol, we train a classifier $f$ to distinguish samples $x$ generated by $g$ and samples drawn from the truth data probability distribution, $p_{\mathcal{X}}$. If the generative model $g$ is part of a GAN, it is possible to use the discriminator $d$ as the classifier~\cite{che2020gan}, but it might be more appropriate to train a new unbiased network from scratch~\cite{2009.03796}. If this classifier is trained with the binary cross entropy (BCE) and converges to its optimum, it obeys the well-known (see e.g. Refs.~\cite{hastie01statisticallearning,sugiyama_suzuki_kanamori_2012}) property
\begin{align}
    f(x)&\approx\frac{p_{\mathcal{X}}}{p_{g(\mathcal{Z})}(x)+p_{\mathcal{X}}}\,.
\end{align}
This can be used to assign a weight
\begin{align}
    w(x)=\frac{f(x)}{1-f(x)}\approx \frac{p_{\mathcal{X}}}{p_{g(\mathcal{Z})}(x)}\,,
\end{align}
to each generated sample.  This idea has been widely studied in HEP for simulation reweighting and inference (see e.g. Ref.~\cite{Martschei_2012,Cranmer:2015bka,Rogozhnikov:2016bdp,Aaij:2017awb,Aaboud:2018htj,Brehmer:2018kdj,Brehmer:2018eca,Brehmer:2018hga,Andreassen:2018apy,Andreassen:2019nnm,Andreassen:2019cjw,Bothmann2019,Andreassen:2020nkr}) and is the core of the \dctr protocol~\cite{Andreassen:2019nnm} that has also been proposed for simulation surrogate refinement~\cite{2009.03796} so we will refer to events weighted in this way as \dctr.

If weighted events are acceptable, then one could stop here and use the weights in the data space (as was proposed in Ref.~\cite{2009.03796}).  For \laser, we next backpropagate this weight to the corresponding latent space point $g(z_i)=x_i$. 
If $g$ is not bijective, then the weights pulled back from the data space to the latent space can be made a proper function of the phase space via Step 2 of the \textsc{OmniFold} algorithm~\cite{Winterhalder:2021ave,Andreassen:2019cjw}. Finally, we sample from the new weighted latent space $(z,\bar{w}(z))$. There are three possible ways proposed by the \laser protocol:
\begin{itemize}
    \item \textbf{Rejection sampling:}
    The weights $w(z)$ can be used to perform rejection sampling on the latent space. In general, however, the weights are broadly distributed which makes this method inefficient and least favored.
    \item \textbf{Markov chain Monte Carlo:} 
    The weights $\bar{w}(z)$ induce a modified probability distribution
    \begin{equation}
        q(z,\bar{w})= p_\mathcal{Z}(z)\,\bar{w}(z)/Z_0,
        \label{eq:mcmc_distribution}
    \end{equation}
    with some normalization constant $Z_0$. If both the probability $q$ and its derivative $\partial q/\partial z$ are tractable, we can employ a Hamiltonian Monte Carlo (HMC) \cite{DUANE1987216,neal2012mcmc} algorithm for sampling.  Differentiability is natural when $w$ is a neural network.
    \item \textbf{Refiner network $\Phi$:}
    Finally, if the derivative $\partial q/\partial z$ is not tractable the weighted latent space $(z,w(z))$ can instead be converted into a unweighted latent space. For this, another generative model $\Phi$ is trained on the weighted latent space~\cite{Backes:2020vka, Verheyen:2020bjw} while generating unweighted samples following the same probability distribution $p_\Phi(z)\approx q(z,\bar{w})$. The output of the refinement network $\Phi$ can then be used as an input to the generative model $g$.
\end{itemize}
In contrast to the MCMC algorithm, the refiner network $\Phi$ requires the training of an additional generative model which might be affected by the same topological obstructions or even evokes new type of problems. Therefore, when applying the \laser protocol in our experiments, we always use a HMC algorithm to sample from the refined latent space when possible.

\subsection{Augmented normalizing flows}
\label{sec:aug_flow}

Augmented normalizing flows~\cite{dupont2019augmented,huang2020augmented,chen2020vflow,nielsen2020survae} are bridging the gap between VAEs and NFs. While they allow for (partially) bijective mappings they also tackle the bottleneck problem~\cite{chen2020vflow} by increasing the dimensionality of the original data. As it has been pointed out in great detail in Ref.~\cite{nielsen2020survae}, augmented flows only represent a particular example of surjective normalizing flows. In general, there are multiple and potentially interesting surjective transformations that might enhance the overall performance of the network~\cite{Verheyen:2022tov}. In our studies, we restrict ourselves to only investigate the effect of an augmentation transformation. As all necessary formulae for surjective transformations have been derived and documented exhaustively in the literature, we only explain briefly how the augmentation transformation is defined and refer for details to Ref.~\cite{nielsen2020survae}.

Augmentation (or slicing in the inverse direction) starts by augmenting the data $x$ with an additional $D_k$ dimensional random variable $z_2\in\mathbb{R}^{D_k}$ which is drawn from a distribution $q(z_2)$.  While not necesary in our application, the more general case allows for $z_2$ to be conditional on the data $x$. In all our examples, $q(z_2)$ is simply given by a normal distribution, \ie $q(z_2)=\mathcal{N}(0,1)$. This means that the forward and the inverse passes are given by
\begin{align}
    x  
    \quad 
    &\stackrel[\leftarrow \text{slice}]{\text{augment} \rightarrow}{\xleftrightarrow{\hspace*{1.5cm}}}
    \quad
    (z_1,z_2),\notag\\
    &\mwith \quad z_1=x, \quad \mand \quad z_2\sim\mathcal{N}(0,1).
\end{align}
As this mapping is surjective in the inverse direction but stochastic in forward direction, the full likelihood of this transformation is not tractable. Consequently, the augmented flow is optimized using only the tractable likelihood contribution which is the ELBO, and connects surjective flows with VAEs.

We note that a similar idea was also introduced in Ref.~\cite{bellagente:2020piv} where a noise-extended INN was used to match the dimensions of parton-level and detector-level events to better describe the stochastic nature of detector effects.

\section{Data representation}
\label{sec:data_representation}

While it is well known that proper data preprocessing can increase the accuracy and reliability of a neural network, it is often difficult to find the best data representation for any particular dataset.

In order to better understand which data representations are well suited in the context of event generation, we investigate several different proposals from the literature and develop our own parametrizations, which are built to enhance the sensitivity of the neural network dependent on its specific task. In detail, we consider parton-level events from a hadronic $2\to n$ scattering process with an intrinsic dimensionality of $d=3n-2$.
We implemented the following preprocessings:
\begin{enumerate}
    \item \textbf{4-Mom}: We parametrize each final state particle in terms of its four momentum and represent the entire event by $4n$ features
    \begin{align}
        x = \{p_1, p_2,\dots, p_n\}
        \label{eq:param_4mom}
    \end{align}
    where $n_f$ denotes the number of final-state particles,
    and standardize each feature according to
    \begin{align}
        \tilde{x}_i = \frac{x_i - \mu_i}{\sigma_i}.
        \label{eq:normalization}
    \end{align}
    
    
    
    \item \textbf{MinRep}: We start by considering the 3-momenta of all final state particles, neglecting the energy component due to on-shell conditions. Further, in order to obey momentum conservation in the transverse plane, we reject the $p_x$ and $p_y$ components of the last particle and treat them as dependent quantities. This reduces the dataset to a $3n-2$ dimensional data representation. Again, we normalize all features using the standardization in Eq.~\eqref{eq:normalization}.
    
    \item \textbf{\prep}: In this preprocessing, we follow the preprocessing suggested in Ref.~\cite{Butter:2021csz} and thus denote it as \textit{precision enthusiast} (\prep) parameterization. For parton level events, this preprocessing starts by representing each finals state in terms of
    \begin{align}
        \{p_\text{T},\eta,\phi\},
    \end{align}
    followed by the additional transformations
    \begin{align}
        p_\text{T}&\to \tilde{p}_\text{T} = \log(p_\text{T}-p_{\text{T},\text{min}}),\\
        \phi_{\Pj_i}&\to \widetilde{\phi}_{\Pj_i}=\atanh(\Delta\phi_{\PW\Pj_i}/\pi),\\
        \phi_{\PW}&\to \widetilde{\phi}_{\PW}=\atanh(\phi_{\PW}/\pi),
    \end{align}
    to gaussanize all features. This leads to $3n$\-/dimensional representation. 
    
    \item \textbf{\mahambo}: In this preprocessing, we map the phase space onto a unit-hypercube with $3n-2$ dimensions. Hence, we reduce the representation to the relevant degrees of freedom similar to the MinRep approach. While the MinRep parametrization is dependent on the ordering of the momenta and hence potentially prone to instabilities, the \mahambo preprocessing relies on the \rambo~\cite{Rambo, Platzer:2013esa} algorithm which populates the phase space uniformly (at least for massless final states). The details about the \mahambo algorithm are given in more detail in Sec.~\ref{sec:mahambo}.
    
    \item \textbf{\elfs}: This starts from a \prep inspired representation
    \begin{align}
        \{\log p_\text{T}, \widetilde{\phi}, \eta\}
    \end{align}
    and additionally augments with the features
    \begin{align}
        \{\atanh(2\Delta\phi_{ij}/\pi-1), \log(\Delta\eta_{ij}),
        \log(\Delta R_{ij})\}
    \end{align}
    for all possible jets pairs $ij$, where we define
    \begin{align}
        \Delta \eta_{ij} &= \vert \eta_i - \eta_j \vert\notag\\
        \Delta \phi_{ij} &= 
        \begin{cases}
            \vert \phi_i - \phi_j \vert, & \text{if}\, \vert \phi_i - \phi_j \vert < \pi \\
            2\pi - \vert \phi_i - \phi_j \vert, & \text{otherwise}
        \end{cases}\notag\\
        \Delta R_{ij} &= \sqrt{(\Delta \phi_{ij})^2 + (\Delta \eta_{ij})^2}.
    \end{align}
    This yields a $3n+k$ dimensional representation, where $k=\{3,9\}$ for 2- or 3-jet events considered in our studies, respectively. Thus, this is denoted as \textit{enlarged feature space} (\elfs) representation.
\end{enumerate}
The list of possible parametrizations above is not exhaustive, but they cover common representations and thus we expect them to be representative. 

\subsection{The \mahambo algorithm}
\label{sec:mahambo}

In order to reduce the dataset to the relevant degrees of freedom while avoiding the limitations of the MinRep parametrization, which is sensitive to the order of the particle momenta, we employ a modified version of the \rambo~\cite{Rambo, Platzer:2013esa} algorithm which allows for an invertible mapping between momenta $p$ and random numbers $r$ living on the unit-hypercube $U=[0,1]^d$.
While the original \rambo~\cite{Rambo} algorithm is not invertible as it requires $4n$ random numbers to parametrize $n$ particle momenta, the \rambodiet~\cite{Platzer:2013esa} algorithm solves this problem and only requires $3n-4$ random numbers. By construction, the \rambodiet algorithm is defined for a fixed partonic center-of-mass (COM) energy and provides momenta within this frame.

However, as we are considering scattering processes at hadron colliders the corresponding events only obey momentum conservation in the transverse plane and are only observed in the lab frame. Consequently, we need two additional random numbers $r_{3n-3}$, $r_{3n-2}$ that parametrize the momentum fractions $x_1$ and $x_2$ of the scattering partons, which are relevant for the PDFs and connect the lab frame with the partonic COM frame via a boost along the $z$-axis. This boost can be parametrized in terms of a rapidity parameter
\begin{align}
    \zeta = \frac{1}{2}\log\frac{x_1}{x_2}=\atanh\frac{q^3}{q^0},\quad \mwith \quad q=\sum_i p_i.
    \label{eq:boosting}
\end{align}
In particular, we choose the following parametrization
\begin{align}
    \tau = \tau_\text{min}^{r_{3n-3}},\quad x_1=\tau^{r_{3n-2}},\quad x_2=\tau^{1-r_{3n-2}},
    \label{eq:lumi}
\end{align}
with $\tau_\text{min}$ reflecting a possible cut on the squared partonic center-of-mass energy $\hat{s}>\hat{s}_\text{min}=\tau_\text{min}s$. We denote the modified algorithm as \mahambo.

The derivation and proofs for the unmodified parts adopted from \rambodiet are given in Ref.~\cite{Platzer:2013esa}\footnote{Note that there is an error in the original paper in the transformation from Eq.\,(8) to Eq.\,(9). Thus, the algorithm has some missing squares in the algorithm when solving for the variable $u_i$. These are fixed in our implementation.}. 
%
\begin{algorithm}[!htbp]
    \caption{The \mahambo algorithm.}
    \begin{algorithmic}[1]
        \STATE $\tau\leftarrow\tau_\text{min}^{r_{3n-3}}$, $x_1\leftarrow\tau^{r_{3n-2}}$, $x_2\leftarrow\tau^{1-r_{3n-2}}$, $\mathcal{W}\leftarrow\sqrt{\tau s}$
        \STATE $Q\leftarrow (\mathcal{W},0,0,0)$, $Q_1\leftarrow Q$, $M_1 \leftarrow \sqrt{Q^2}$, $M_n\leftarrow0$
        \FOR{$i=2,\dots,n$}
        \IF{$i\ne n$}
        \STATE solve \\
        $r_{i-1}=(n+1-i)u_i^{2(n-i)}-(n-i)u_i^{2(n+1-i)}$ \\for $u_i$
        \STATE $M_i\leftarrow u_2\dots u_i \sqrt{Q^2}$
        \ENDIF
        \STATE $\cos\theta_i\leftarrow 2r_{n-5+2i}-1$, $\sin\theta_i\leftarrow\sqrt{1-\cos^2\theta_i}$, $\phi_i\leftarrow 2\pi r_{n-4+2i}$
        \STATE $q_i\leftarrow 4M_{i-1}\rho(M_{i-1},M_i,0)$
        \STATE $\mathbf{p}_{i-1}\leftarrow q_i (\cos\phi_i\sin\theta_1, \sin\phi_i\sin\theta_i, \cos\theta_i)$
        \STATE $p_{i-1}\leftarrow(q_i,-\mathbf{p}_{i-1})$, $Q_i\leftarrow(\sqrt{q_i^2+M_i^2},-\mathbf{p}_{i-1})$
        \STATE boost $p_{i-1}$ and $Q_i$ by $\mathbf{Q}_{i-1}/Q_{i-1}^0$
        \ENDFOR
        \STATE $p_n\leftarrow Q_n$
        \STATE solve $\mathcal{W}=\sum_{i=1}^n\sqrt{\xi^2\!\left(p_i^0\right)^2 + m_i^2}$ for $\xi$
        \FOR{$i=1,\dots,n$}
        \STATE $p_i^0\leftarrow\sqrt{\xi^2\!\left(p_i^0\right)^2 + m_i^2}$, $\mathbf{p}_i\leftarrow \xi\mathbf{p}_i$
        \STATE boost $p_i$ by $\Lambda_z(\zeta)$ with rapidity $\zeta\leftarrow\frac{1}{2}\log\frac{x_1}{x_2}$
        \ENDFOR
        \RETURN $\{p_1,\dots,p_n\}$
    \end{algorithmic}
    \label{al:mahambo}
\end{algorithm}
%
The \mahambo procedure to generate a single phase-space $x$ point starting from random numbers $r$ passes the following steps:
\begin{enumerate}
    \item Sample the partonic momentum fractions $x_1$ and $x_2$ according to Eq.~\eqref{eq:lumi}.
    \item Define the partonic COM energy $\hat{s}=\sqrt{x_1 x_2 s}$ and define the Lorentz boost parameter $\zeta$ according to Eq.~\eqref{eq:boosting}.
    \item Pass $\hat{s}$ to the \rambodiet algorithm to generate $n$ massless 4-momenta in the partonic COM frame.
    \item Possibly reshuffle these four\-/momenta following the procedure in Ref.~\cite{Rambo} to obtain massive final states.
    \item Boost all momenta into the lab frame with $\Lambda_z(\zeta)$.
\end{enumerate}
The full details of the mapping between $3n-2$ random numbers $r_i$ and $n$ 4-momenta is given in Algorithm~\ref{al:mahambo}, and its inverse in Algorithm~\ref{al:inv_mahambo}. 
%
\begin{algorithm}[!htbp]
    \caption{The inverse \mahambo algorithm.}
    \begin{algorithmic}[1]
        \STATE $q\leftarrow\sum_{j=1}^{n}p_j$, $\tau\leftarrow q^2/s$, $\zeta\leftarrow \atanh(q^3/q^0)$
        \STATE $\mathcal{W}\leftarrow\sqrt{\tau s}$, $\log x_1\leftarrow\frac{1}{2}\left[\log\tau + \log \left(\frac{1-\zeta}{1+\zeta}\right)\right]$
        \STATE $r_{3n-3}\leftarrow \log\tau/\log\tau_\text{min}$, $r_{3n-2}\leftarrow \log x_1/\log\tau$
        \STATE solve $\mathcal{W}=\sum_{i=1}^n\sqrt{\left(\left(p_i^0\right)^2 - m_i^2\right)/\xi^2}$ for $\xi$
        \FOR{$i=1,\dots,n$}
        \STATE $p_i^0\leftarrow\sqrt{\left(\left(p_i^0\right)^2 - m_i^2\right)/\xi^2}$, $\mathbf{p}_i\leftarrow \mathbf{p}_i/\xi$
        \STATE boost $p_i$ by $\Lambda_z(-\zeta)$ with rapidity $\zeta$
        \ENDFOR
        \STATE $M_{1}\leftarrow\sqrt{\left(\sum_{j=1}^{n}p_j\right)^2}$, $Q_n\leftarrow p_n$
        \FOR{$i=n,\dots,2$}
        \STATE $M_{i}\leftarrow\sqrt{\left(\sum_{j=i}^{n}p_j\right)^2}$
        \IF{$i\ne n$}
        \STATE $u_i\leftarrow M_i/M_{i-1}$
        \STATE $r_{i-1}\leftarrow(n+1-i)u_i^{2(n-i)}-(n-i)u_i^{2(n+1-i)}$
        \ENDIF
        \STATE $Q_{i-1}\leftarrow Q_i+p_{i-1}$
        \STATE boost $p_{i-1}$ by $-\mathbf{Q}_{i-1}/Q_{i-1}^0$
        \STATE $r_{n-5+2i}\leftarrow(p_{i-1}^3/\vert\mathbf{p}_{i-1}\vert+1)/2$
        \STATE $\phi\leftarrow\arctan(p_{i-1}^2/p_{i-1}^1)$, $r_{n-4+2i}\leftarrow\phi/2\pi +\Theta(-\phi)$
        \ENDFOR
        \RETURN $\{r_1,\dots,r_{3n-2}\}$
    \end{algorithmic}
    \label{al:inv_mahambo}
\end{algorithm}
%
To solve the equations in line 5 and 15 of Algorithm~\ref{al:mahambo} and line 4 of Algorithm~\ref{al:inv_mahambo} we use Brent's method~\cite{brent-method}. The function $\rho(M_{i-1},M_i,m_{i-1})$ in line 9 corresponds to the two-body decay factor defined as
\begin{align}
    \rho(a,b,c)
    =\frac{\sqrt{(a^2-(b+c)^2)(a^2-(b-c)^2)}}{8a^2},
\end{align}
which simplifies in the massless ($m_i=0$) case to
\begin{align}
    \rho(M_{i-1},M_i,0)=\dfrac{M_{i-1}^2-M_{i}^2}{8M_{i-1}^2}.
\end{align}
%

\section{Experiments}
\label{sec:experiments}

We demonstrate the performance of \laser and augmented flows in two numerical experiments. We start with two illustrative toy examples in Section~\ref{sec:toy_examples}, before turning to a particle physics problem in Section~\ref{sec:hep_example}. For the more complicated examples, we further perform a dedicated analyses of choosing an optimal parametrization for both the generative model as well as the classifier needed for the \dctr or \laser approach.

In order to have full control of all parameters and inputs we implemented our own version of Hamiltonian Monte Carlo in \pytorch, using its automatic differentiation module to compute the necessary gradients to run the algorithm. For all examples, we initialize a total of 100 Markov Chains running in parallel to reduce computational overhead and solve the equation of motions numerically using the leapfrog algorithm~\cite{neal2012mcmc} with a step size of $\epsilon = 0.01$ and 30 leapfrog steps. To avoid artifacts originating from the random initialization of the chains, we start with a burn-in phase and discard the first 5000 points from each chain.

\subsection{Toy examples}
\label{sec:toy_examples}

We consider two different toy examples commonly used in the literature: a probability density defined by thin concentric circles and another density described by circular array of two-dimensional Gaussian random variables.  These are illustrated pictorially in in the leftmost column of Fig.~\ref{fig:toy_examples}.

\subsubsection*{Network architecture}

We use a normalizing flow with affine coupling blocks as the baseline model in all toy examples. Our implementation relies partially on \survae~\cite{nielsen2020survae}. The NF consists of 8 coupling blocks, each consisting of a fully connected multi-layer perceptron (MLP) with two hidden layers, 32 units per layer, and ReLU activation functions. The model is trained for 100 epochs by minimizing the negative log-likelihood with an additional weight decay of $10^{-5}$. The optimization is performed using Adam~\cite{Adam2019} with default $\beta$ parameters and $\alpha_0 = 10^{-3}$. We employ an exponential learning rate schedule, \ie $\alpha_n = \alpha_0 \gamma^{n}$ , with $\gamma = 0.995$ which decays the learning rate after each epoch.

In the scenario using the augmented flow, we extend the dimensionality from $2\to 4$ by augmenting an additional 2-dimensional random number $r\sim\mathcal{N}(0,1)$ to the feature vector $x$.

For the classifier, we use a simple feed-forward neural network which consists of 8 layers with 80 units each and leaky ReLU activation~\cite{Maas2013RectifierNI} in the hidden layers. 
We note that, none of these hyperparameters have been extensively optimized in our studies.

\subsubsection*{Probability distance measures}

In the 2-dimensional toy examples, we can quantify the performance gain of our \laser-flow and the augmented flow compared to our baseline flow model. For this, we implemented different $f$-divergences~\cite{2014}. All $f$-divergences used in this paper are shown in Tab.~\ref{tab:f-divergencies}.
\begin{table*}[t]
    \centering
    \begin{tabular}{llc}
        \toprule
        \parbox{3cm}{\centering Name} & \parbox{5cm}{\centering Formula $f(p,q)$} & \parbox{3cm}{\centering Proper metric} \\ 
        \midrule
        Total variation (V) & $\frac{1}{2}\int\d x\,\left\vert p(x)-q(x)\right\vert$  &  {\color{green!70!black}\cmark} \\
        Squared Hellinger (H$^2$) & $\frac{1}{2}\int \d x\,\left(\sqrt{p(x)}-\sqrt{q(x)}\right)^2$ & {\color{red!80!black}\xmark}  \\
        Jensen-Shannon (JS) & $\frac{1}{2}\int \d x\,p(x)\log\!\left(\frac{2p(x)}{p(x)+q(x)}\right)+(p\leftrightarrow q)$ & {\color{red!80!black}\xmark}  \\
        \midrule
        \parbox{3cm}{\centering Earth mover distance (EMD)}& \parbox{5cm}{\centering Eq.~\eqref{eq:emd}} & {\color{green!70!black}\cmark}  \\
        \bottomrule
    \end{tabular}
    \caption{A comparison of different $f$-divergences and the earth mover distance (EMD).}
    \label{tab:f-divergencies}
\end{table*}
Except for the total variation (V), the squared Hellinger distance (H$^2$) and the Jensen-Shannon divergence (JS) are not proper metrics as they do not obey the triangle-inequality. In contrast to the Kullback–Leibler (KL) divergence, which is also commonly used, we only considered $f$-divergences which are symmetric in $p$ and $q$.

Additionally, we implemented the earth mover's distance (EMD)~\cite{emd} between two probability distributions $p(x)$ and $q(x)$. The EMD is directly connected to an optimal transport problem. If the two distributions are represented as a set of tuples $P=\{(x_i,p(x_i)\}_{i=1}^N$ and $Q=\{(y_j,q(y_j)\}_{j=1}^{M}$, the EMD is the minimum cost required to transform $P$ into $Q$. This transformation is parameterized by discrete probability flows $f_{ij}$ from  $x_i\in P$ to $y_j\in Q$ and reads
\begin{align}
    \begin{split}
        \mathrm{EMD}(p,q)&=\min_{\{f_{ij}>0\}}\sum_{i=1}^N \sum_{i=1}^{M}\,f_{ij}\,||x_i - y_j||_2, \\
        \sum_{j=1}^{M}\,f_{ij}\leq &p(x_i),\quad \sum_{i=1}^{N}\,f_{ij}\leq q(y_j),\\
        \sum_{i=1}^N \sum_{i=1}^{M}\,f_{ij}&=\min\!\left(\sum_{i=1}^{N}p(x_i), \sum_{j=1}^{M}q(y_j)\right),
    \end{split}
    \label{eq:emd}
\end{align}
where $\vert\vert\cdot\vert\vert_2$ is the euclidean norm and the optimal flow $f_{ij}$ is found by solving the optimization problem. For an efficient numerical implementation we use the \textsc{POT} library~\cite{flamary2021pot}. An advantage of the EMD over the $f$-diver\-gences is that it also takes the metric space of the points $x_i$ and $y_i$ into account.

All $f$-divergences introduced have been evaluated in a histogram-based way using 1M data points for all models and examples. By comparing the same metrics for two equally large, but independent samples drawn from the truth distribution, we can estimate the uncertainty of the quantities. The extracted error estimates are indicated in parenthesis in Table~\ref{tab:scores}. 

We note that we used a total of $10^4=100\times 100$ bins to provide a detailed test of the phase space for evaluating all metrics while still providing stable results for the given statistics. We did not optimize the number of bins, but found that the qualitative conclusions are unaffected by the details of this choice.

\subsubsection*{Results}

\begin{figure*}[!htbp]
    \centering
    \includegraphics[width=0.85\textwidth]{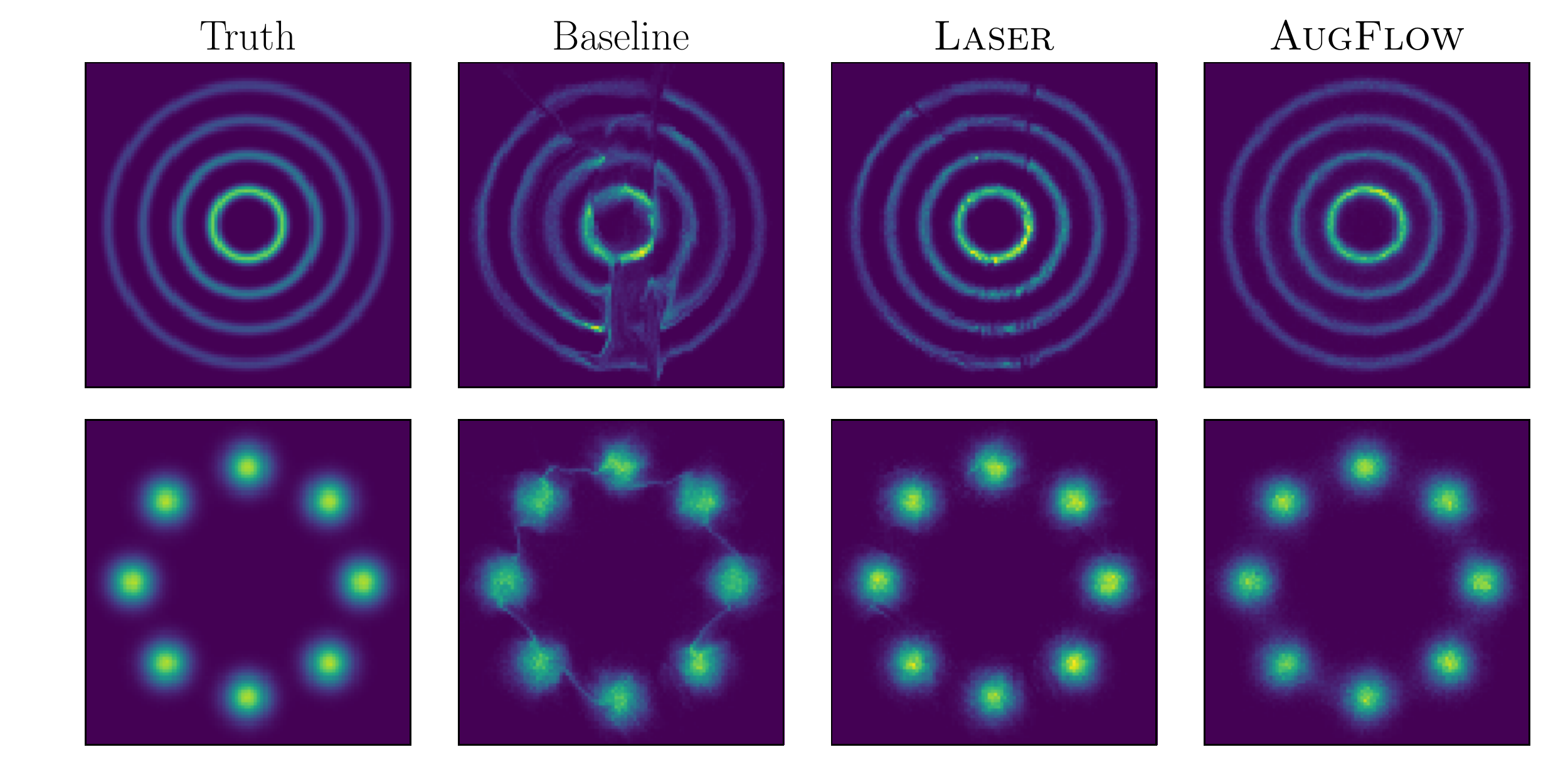}
    \caption{Comparison of the baseline flow model, the \textsc{LaSeR} refined output and the augmented flow for the Circle (top) and Gaussians (bottom) toy examples.}
    \label{fig:toy_examples}
\end{figure*}

In the first toy example, illustrated in the top panel of  Fig.~\ref{fig:toy_examples}, we considered multiple concentric circles representing a multimodal distribution that is topologically different than the simple latent space given by a Gaussian random variable.
Consequently, the baseline flow fails to properly learn the truth density and generates fake samples which result in a blurred distribution that is not able to properly separate the modes. In contrast, the \laser improved generator as well as the \augflow show highly improved distributions which is both easily detectable by eye as well as quantified by the lower divergences given in Table~\ref{tab:scores}. Overall, when looking at the given scores in Table~\ref{tab:scores}, the performance of the \laser improved model is superior over both the baseline model as well as the \augflow, except for the EMD metric where the \augflow yields the best score.

\begin{table}[hbtp!]
    \centering
    \begin{tabular}{c|lcccc}
        \toprule
        \multicolumn{2}{c}{Method}&  EMD  & V & H$^2$ & JS \\
        \midrule
        \multirow{3}*{Circle} & Baseline 
        & $0.0153(2)$  & $0.24(3)$  & $0.069(1)$  & $0.0610(9)$ \\
        & \laser 
        & $0.0075(2)$ & $\mathbf{0.10(3)}$ & $\mathbf{0.012(1)}$  & $\mathbf{0.0115(9)}$  \\
        & \augflow  
        & $\mathbf{0.0041(2)}$ & $0.12(3)$ & $0.024(1)$  & $0.0211(9)$ \\
        \midrule
        \multirow{3}*{Gaussians} & Baseline 
        & $0.0048(9)$  & $0.10(3)$  & $0.013(1)$  & $0.012(1)$ \\
        & \laser 
        & $\mathbf{0.0035(9)}$ & $\mathbf{0.05(3)}$ & $\mathbf{0.005(1)}$  & $\mathbf{0.004(1)}$  \\
        & \augflow 
        & $0.0046(9)$ & $0.07(3)$ & $0.007(1)$  & $0.007(1)$ \\
        \bottomrule
    \end{tabular}
    \caption{Earth mover distance (EMD), total variation (V), squared Hellinger distance (H$^2$) and Jensen–Shannon divergence (JS) on test data for the baseline model and the proposed methods for various two-dimensional examples. The best results are written in bold face. The errors on the scores are indicated in parenthesis.}
    \label{tab:scores}
\end{table}

Similarly to the first example, the baseline model is again not expressive enough to adequately describe the second toy example, illustrated in the lower panel of Fig.~\ref{fig:toy_examples}. Both \laser and \augflow are nearly indistinguishable from the truth by eye, which is quantified by even lower scores numerically than for the circle example in Table~\ref{tab:scores}. In this case, the \laser improved flow gives the best score for all evaluated metrics and outperforms the \augflow. This indicates that the \laser protocol seems to typically perform better than the augmentation of additional dimensions, an observation we will make again in the higher dimensional physics examples in Sec.~\ref{sec:hep_example}.

Finally, we want to remark that simply ramping up the number of parameters in the baseline model would also improve its performance in these simple low\-/dimensional toy examples. However, this parameter growth scales poorly for more complicated and higher\-/dimensional distributions, leading to increasingly minor improvements. Further, as Ref.~\cite{Winterhalder:2021ave} noted, making a bijective generative model larger will not fundamentally overcome the topology problem.

\subsection{Physics example}
\label{sec:hep_example}

As a representative HEP example, we consider the
\begin{align}
    \Pp \Pp \to \PWp + \{2,3\}\jets
\end{align}
scattering process. For all simulations in this paper, we consider a centre-of-mass energy of $\sqrt{s}= 14\, \mathrm{TeV}$. All events used as training and test data are generated at parton level using \madgraph~\cite{Alwall:2014hca} (v3.4.1). We use the NNPDF 3.0 NLO PDF set \cite{NNPDF:2014otw} with a strong coupling constant $\alpha_s(M_\mathrm{Z}) = 0.119$ in the four flavor scheme provided by \lhapdf~\cite{Buckley:2014ana}. In order to stay in the perturbative regime of QCD we apply a cut on the transverse momenta of the jets as
\begin{align}
    p_{\mathrm{T},\mathrm{j}}>20\,\mathrm{GeV}.
\end{align}
Additionally, we require the jets to fulfill
\begin{align}
    \vert \eta_{\mathrm{j}}\vert<6\,,\qquad \Delta R_{\mathrm{j}\mathrm{j}}>0.4.
\end{align}

We have 1M data points for each jet multiplicity and train on 800k events each. The network is trained on each multiplicity separately. In general, it is also possible to set up an architecture which can be trained on different multiplicities simultaneously~\cite{Butter:2021csz, Verheyen:2022tov}.

\subsubsection*{Network architecture}

For the physics case considered in this work, the affine coupling blocks are replaced with rational quadratic spline (RQS) blocks~\cite{durkan2019neural} with 10 spline bins unless mentioned otherwise.
In order to cope with the higher dimensions we now use 14 coupling blocks, where each block has 2 hidden layers, 80 units per layer and ReLU activation functions. We now employ a one-cycle training schedule with a starting and maximum learning rate of $\alpha_0 = 5\cdot10^{-4}$ and $\alpha_\text{max} = 3\cdot10^{-3}$, respectively. The networks are trained for 50 epochs.
In contrast to the toy examples, the classifier now consists of 10 layers\footnote{We wanted to be sure in results presented later that the network expressiveness was not a limiting factor.} with 128 units each and all other parameters unchanged. 

A possible bottleneck, when running the \laser protocol in terms of a HMC on the latent space, is the necessary fine tuning of the step size $\epsilon$. Usually, the average acceptance rate of a proposed phase-space point is a good indication about the quality of the chosen hyperperameters. In an optimal scenario, the acceptance rate of an HMC should be around $\sim0.65$~\cite{neal2012mcmc}. Consequently, during the initial burn-in phase, we adapt the step-size to achieve the optimal acceptance rate of $0.65$ on average. We found, that this additional adaptive phase was only important for the physics examples considered.

Further, for the augmented flow, we have a freedom in choosing the dimensionality of the augmented random number $r\in\mathbb{R}^{D_r}$. We find that for both physics examples considered we obtain the best performance when choosing $D_r=d$, where $d$ is the dimension of the used data representation. We note, however, that we did not do any extensive tuning of this parameter. 

\subsubsection*{Two-jet events}

Before diving into the results of the various tested methods, we want to emphasize that the structure and storyline of what is coming next are pedagogical. We will start by using a very naive model to highlight and explain the reason for its limitations. Only then will we move forward by alleviating most of these issues and finish this section with a recommended workflow for multi-dimensional event generation.

\begin{figure*}[htbp!]
    \centering
    \includegraphics[page=1,width=0.32\textwidth]{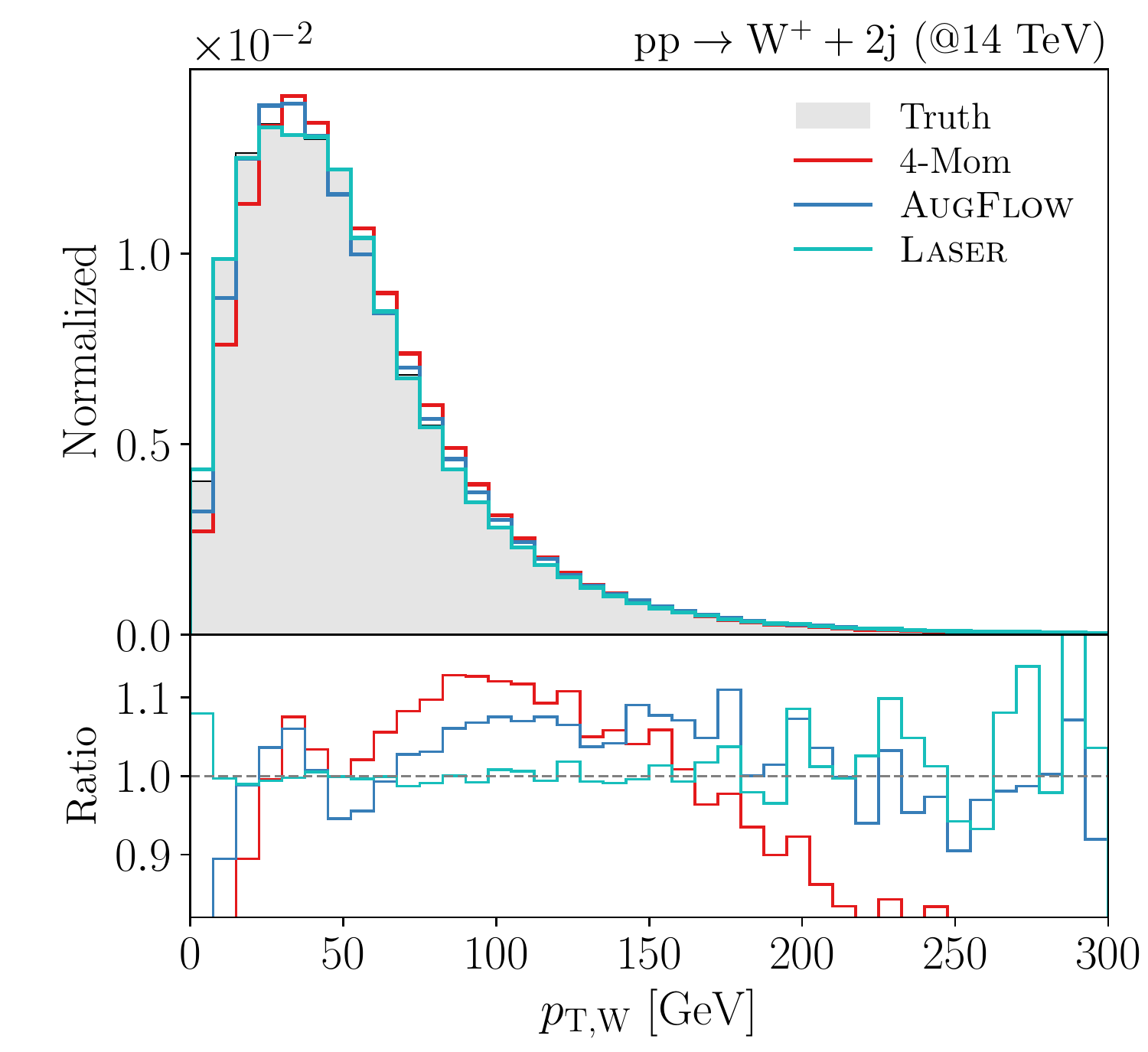}
    \includegraphics[page=3,width=0.32\textwidth]{fighep/lhc/1_wp2j_simple_compr.pdf}
    \includegraphics[page=5,width=0.32\textwidth]{fighep/lhc/1_wp2j_simple_compr.pdf}\\
    \includegraphics[page=7,width=0.32\textwidth]{fighep/lhc/1_wp2j_simple_compr.pdf}
    \includegraphics[page=8,width=0.32\textwidth]{fighep/lhc/1_wp2j_simple_compr.pdf}
    \includegraphics[page=9,width=0.32\textwidth]{fighep/lhc/1_wp2j_simple_compr.pdf}
    \caption{Transverse momentum distributions for the W-boson and the two jets (top), and high-level angular distributions (bottom) for the $\Pp \Pp \to \PWp + 2\Pj$ process. We show the baseline model as well as \textsc{AugFlow} and a \laser improved version. The baseline model uses the 4-vector parametrization introduced in Eq.~\eqref{eq:param_4mom}.}
    \label{fig:wp2j_baseline}
\end{figure*}

\paragraph{Naive baseline model}
In our first study, shown in Fig.~\ref{fig:wp2j_baseline}, we tested a simple affine flow as baseline model which has been trained on $\PWp + 2\Pj$ events given in the naive 4-Mom parametrization. This parametrization represents the most simple and straightforward parametrization possible and consequently serves as a benchmark point for all further improvements. Thus, we also kept the affine blocks in this example to keep the architecture as close to the ones used in the toy examples as possible. Within this parametrization, we train our classifier model which subsequently yields the necessary weights to perform the \laser protocol. As we can see in Fig.~\ref{fig:wp2j_baseline}, both \laser\footnote{The data-space \dctr results are nearly indistinguishable from the \laser results and are therefore omitted.} and the \augflow improve the precision of the baseline network. However, the improvement obtained by the \augflow is 
minimal compared to the large gain in precision by using the \laser protocol. In particular, we find that the more expressive the baseline model is, the less effective the \augflow performance is in these high-dimensional examples. Consequently, we omit showing further comparisons of the \augflow with the baseline model for all other tests performed and focus on techniques that are more relevant for reaching a higher precision.

\paragraph{Different parametrizations in the generator}

\begin{figure*}[htbp!]
    \centering
    \includegraphics[page=1,width=0.32\textwidth]{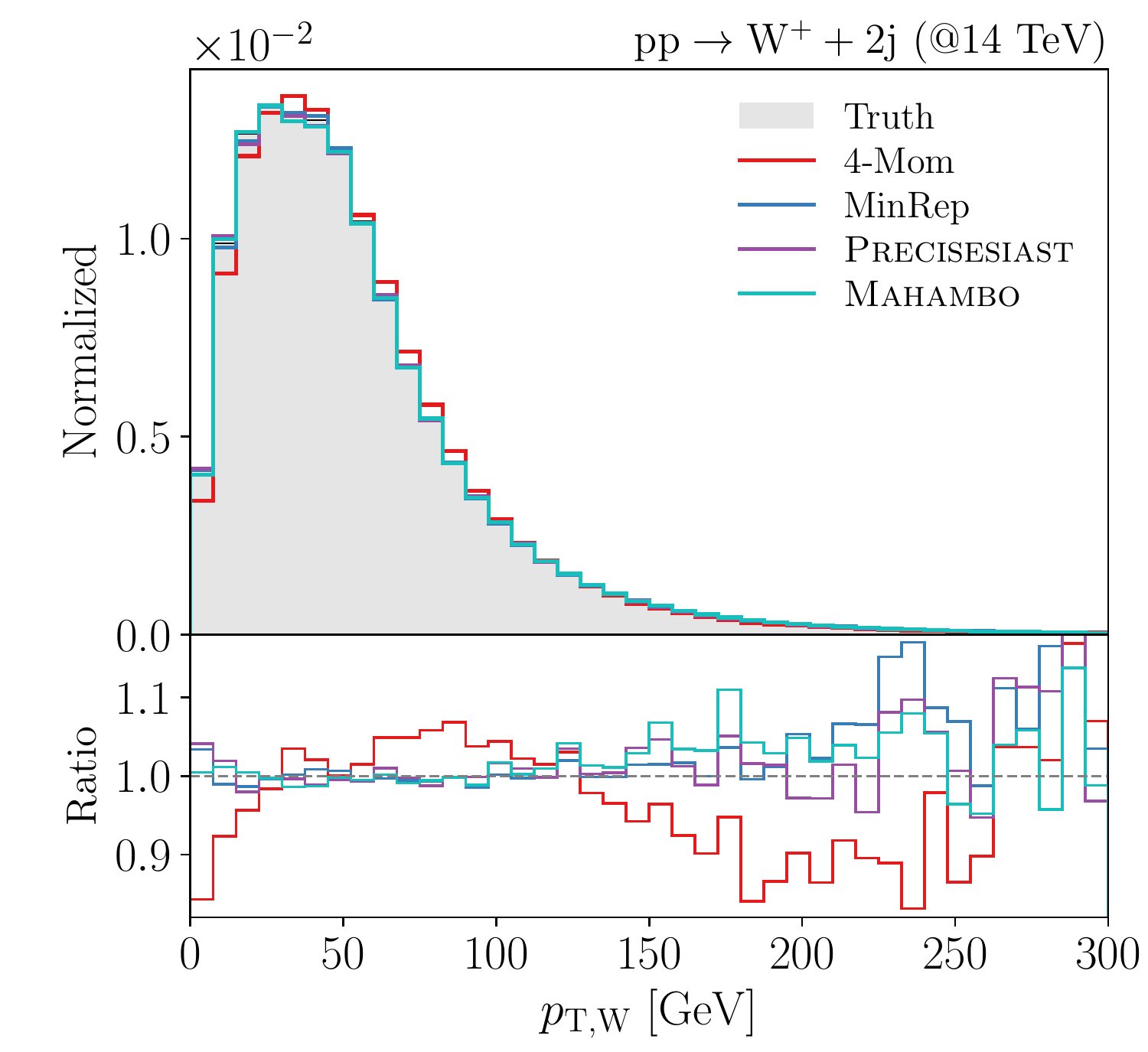}
    \includegraphics[page=3,width=0.32\textwidth]{fighep/lhc/2_wp2j_rqs_comp.pdf}
    \includegraphics[page=5,width=0.32\textwidth]{fighep/lhc/2_wp2j_rqs_comp.pdf}\\
    \includegraphics[page=7,width=0.32\textwidth]{fighep/lhc/2_wp2j_rqs_comp.pdf}
    \includegraphics[page=8,width=0.32\textwidth]{fighep/lhc/2_wp2j_rqs_comp.pdf}
    \includegraphics[page=9,width=0.32\textwidth]{fighep/lhc/2_wp2j_rqs_comp.pdf}
    \caption{Transverse momentum distributions (top) for the final state particles and high-level angular distributions (bottom) for the $\Pp \Pp \to \PWp + 2\Pj$ process. We show the results for different parametrizations used for the generation.}
    \label{fig:wp2j_rqs_param}
\end{figure*}

Next, we study the effect of using different data preprocessings on the quality of the generated events. In particular, we compare the parametrizations 4-Mom, MinRep, \prep and \mahambo which have been introduced in Sec.~\ref{sec:data_representation}. In order to be more compatible with state-of-the-art generators, we now replace the affine blocks with RQS blocks and keep them throughout the rest of our analyses. In the various distributions shown in Fig.~\ref{fig:wp2j_rqs_param}, we see that the previously tested 4-Mom parametrization is strictly the worst, likely because it has the highest dimensionality. This leaves the network with the task of learning the likelihood of a lower-dimensional manifold within this higher\-/dimensional representation, making it harder to match the redundant features. Consequently, the naive MinRep parametrization, which merely omits all redundant features in the data representation, outperforms the 4-Mom preprocessing and achieves higher precision in all observables. 

The results become even better when picking either the \prep or \mahambo preprocessing, as can be seen in Fig.~\ref{fig:wp2j_rqs_param}. At first glance, this seems to somewhat conflict with our conjecture about dimensionality since the \prep parametrization only reduces the data representation to $3n$ dimensions and is thus not minimal. However, this parametrization is deliberately designed to perform well at the otherwise hard-to-learn phase-space boundaries introduced by the $p_{\text{T},\Pj}$ cuts. As a result, this parametrization yields the best precision in all transverse-momentum observables. In contrast, the \prep parametrization fails to correctly describe the $\Delta R_{\Pj\Pj}$ distribution. This is in line with the observations made in Ref.~\cite{Butter:2021csz}, where the authors introduced a `magic' transformation to smooth out the hard phase-space cut.
In comparison, within the \mahambo preprocessing, the flow renders this phase-space boundary automatically and is generally more precise in all connected angular observables such as $\Delta \phi_{\Pj\Pj}$ and $\Delta \eta_{\Pj\Pj}$, shown in the lower panel of Fig.~\ref{fig:wp2j_rqs_param}. While there is indeed still room for some improvement even for the flow using the \prep and \mahambo representation, we postpone the combination of better preprocessing with refinement (via the \laser protocol) to the $\PWp + 3\Pj$ process, as the impact will be more important.

\subsubsection*{Three-jet events}

The three-jet case is much more complex than the two-jet case because the dimensionality of the data manifold is relatively smaller than the data space dimensionality.

However, as the relevant phase space is bigger than in the two-jet case, this allows us to study the scaling behavior of our surrogate model for increasing multiplicities and hence phase-space dimensions. This study is relevant as the expected precision of the surrogate model, which is trained on data following a given underlying distribution, solely scales with the dimension and is not affected by the intrinsic precision of the training data. In other words, given a fixed phase-space dimensionality, the surrogate
model's performance will not degrade by replacing the LO training data with more accurate NLO or even NNLO simulation data. Hence, we only test our methods on easy-to-generate LO data for simplicity and without loss of generality. Finally, we expect the effect effective gain of the surrogate model to be more significant for more complicated processes without losing precision. This has been shown in Ref.~\cite{Danziger:2021eeg} and can be explained as the surrogate model is generally much faster to evaluate than event weights derived from first principles. This speed advantage is becoming even bigger for NLO and NNLO scenarios where the evaluation time of the matrix elements becomes the limiting aspect.

\paragraph{Different parametrizations in the generator reloaded}

\begin{figure*}[htbp!]
    \centering
    \includegraphics[page=1,width=0.32\textwidth]{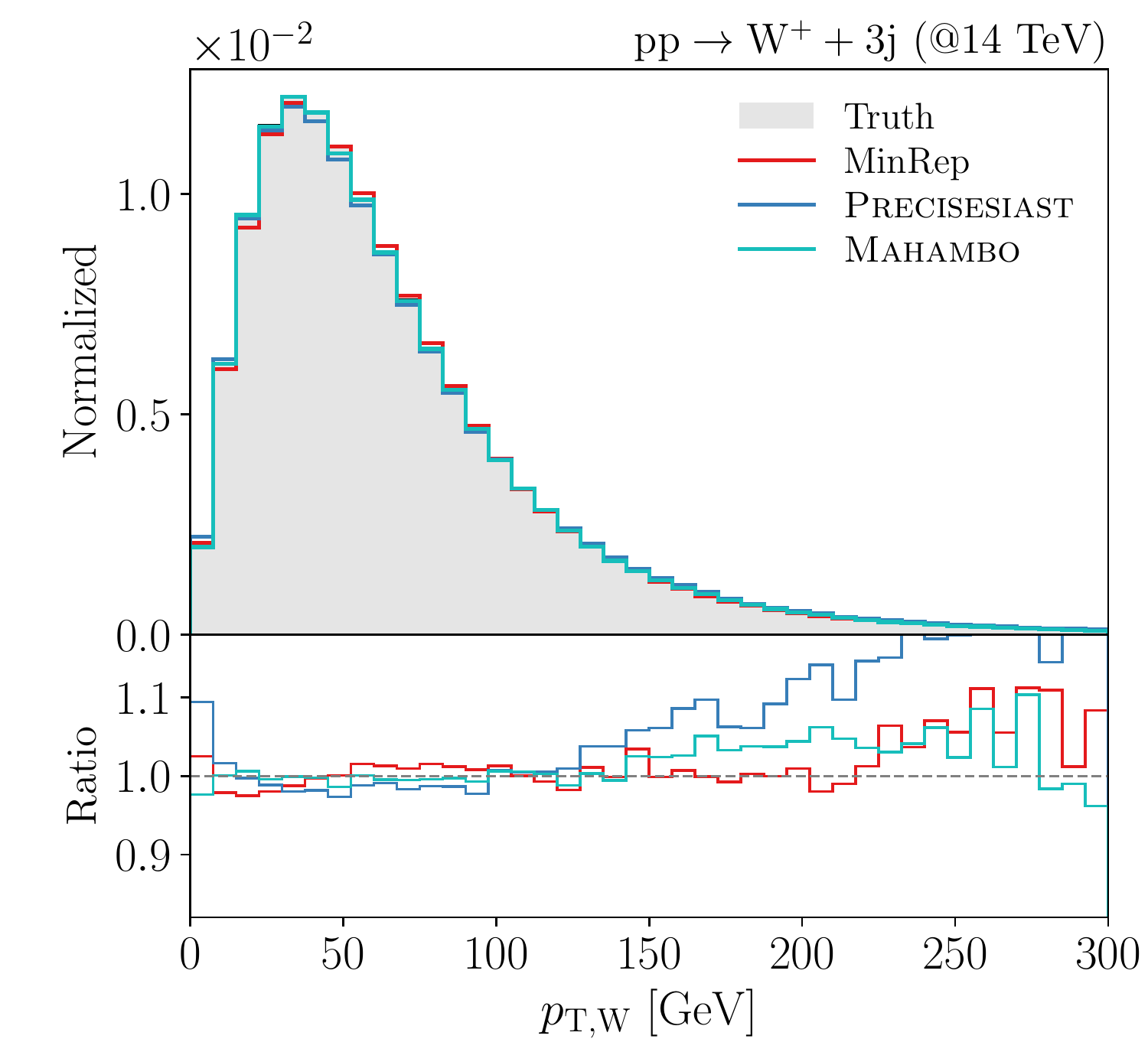}
    \includegraphics[page=3,width=0.32\textwidth]{fighep/lhc/3_wp3j_rqs_comp.pdf}
    \includegraphics[page=5,width=0.32\textwidth]{fighep/lhc/3_wp3j_rqs_comp.pdf}\\
    \includegraphics[page=11,width=0.32\textwidth]{fighep/lhc/3_wp3j_rqs_comp.pdf}
    \includegraphics[page=16,width=0.32\textwidth]{fighep/lhc/3_wp3j_rqs_comp.pdf}
    \includegraphics[page=17,width=0.32\textwidth]{fighep/lhc/3_wp3j_rqs_comp.pdf}
    \caption{Transverse momentum distributions for the W-boson and the two leading jets (top)  and high-level angular diststibutions (bottom) for the $\Pp \Pp \to \PWp + 3\Pj$ process. We show the results for different parametrizations used for the generation.}
    \label{fig:wp3j_rqs_param}
\end{figure*}

First, we study the impact of preprocessing in Fig.~\ref{fig:wp3j_rqs_param} (the three-jet analog of Fig.~\ref{fig:wp2j_rqs_param}).  We omit the 4-Mom parameterization and observe that the MinRep parameterization is already within 10\% across the phase space and often better than or at least as good as \prep and \mahambo.  By design, \prep is excellent for the jet transverse momenta but struggles with some of the $\Delta  R$ observables.  \mahambo is the worse for the second jet transverse momentum but the best at the $\Delta R$ and $\Delta \eta$ between the subleading two jets.  Elsewhere, MinRep and \mahambo are comparable.  

\paragraph{Classifier improved event generation}

The rest of the figures show how reweighting can improve the precision in both the data space and from the latent space.

\begin{figure}[htbp!]
    \centering
    \includegraphics[page=2, width=0.4\textwidth]{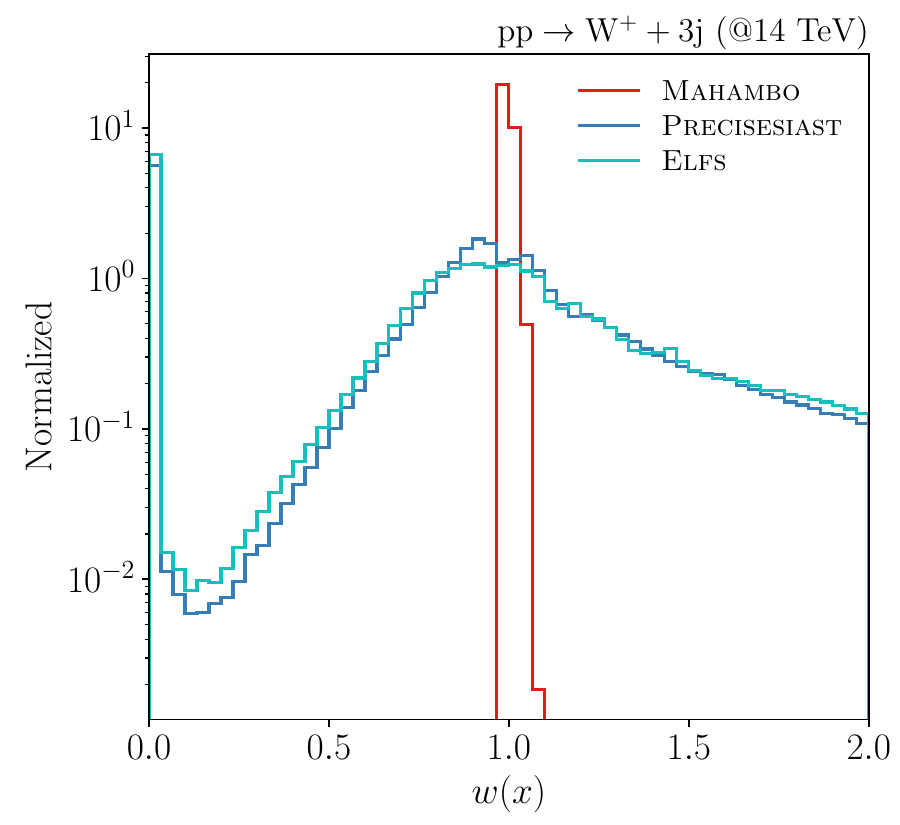}
    \caption{Learned classifier weight $w(x)$ for different parametrizations applied to events generated in the \mahambo parametrization.}
    \label{fig:wp3j_weights}
\end{figure}

Initially, the classifier was fed directly by the events generated in the \mahambo preprocessing, meaning the discriminator acts on a 10\-/dimensional hypercube. However, as shown in Fig.~\ref{fig:wp3j_weights} by the red curve, the classifier is not learning any sensible weights and sharply peaks at $w(x)\approx 1$. This means the classifier cannot distinguish between the generated data and the truth using the \mahambo parametrization even though we tested a rather large network size to ensure that too few network parameters are not limiting the performance. Therefore, to better use the classifier, we change the data parametrization before feeding into the network to enhance the classification capabilities as much as possible. In particular, if the classifier inputs directly contain critical features such as $p_{\text{T},\Pj}$, $\Delta R$, $\Delta \eta$ and $\Delta\phi$ the obtained weights provide much more information, as can be seen by the weight distributions for the \prep and \elfs parametrization shown in Fig.~\ref{fig:wp3j_weights}.

\begin{figure*}[htbp!]
    \centering
    \includegraphics[page=3,width=0.32\textwidth]{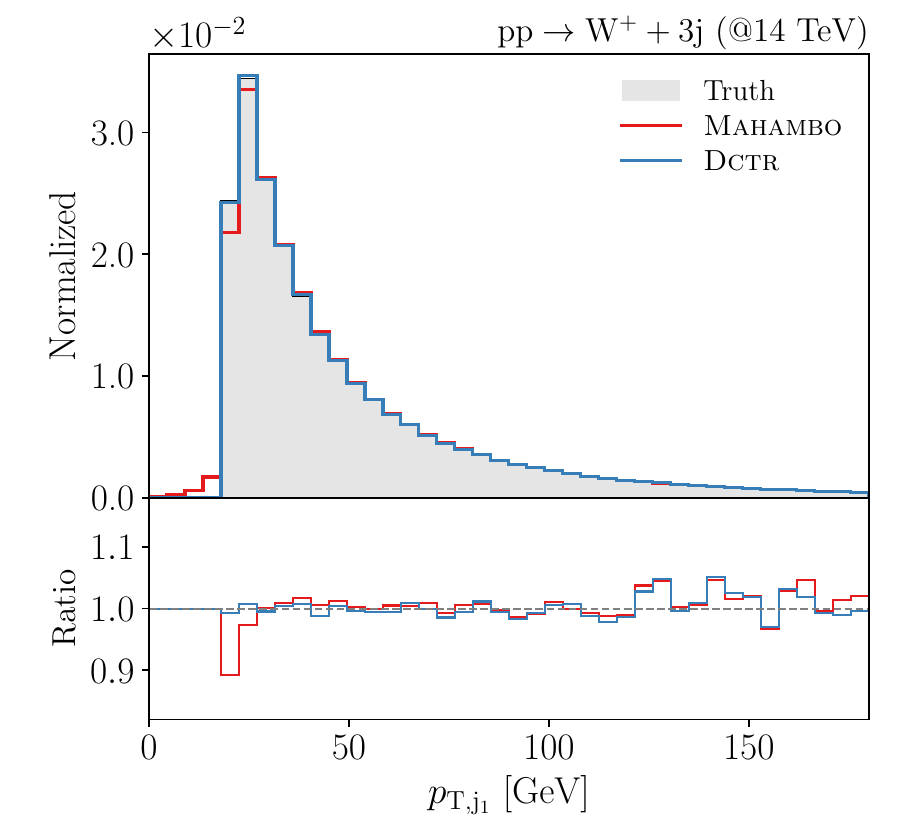}
    \includegraphics[page=9,width=0.32\textwidth]{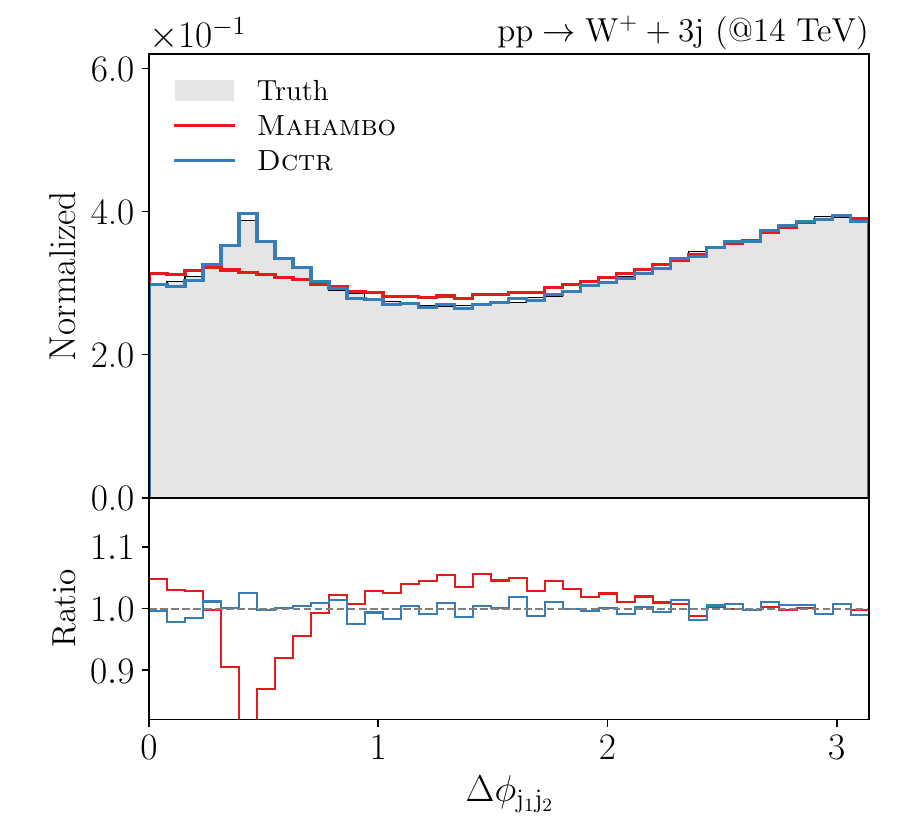}
    \includegraphics[page=11,width=0.32\textwidth]{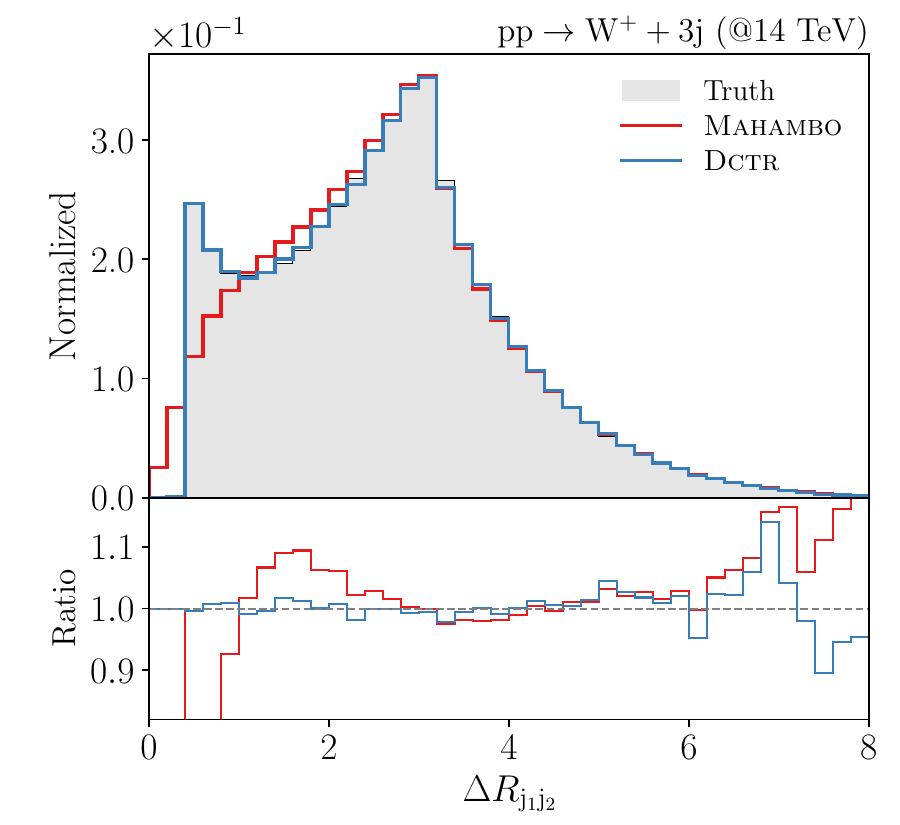}\\
    \includegraphics[page=15,width=0.32\textwidth]{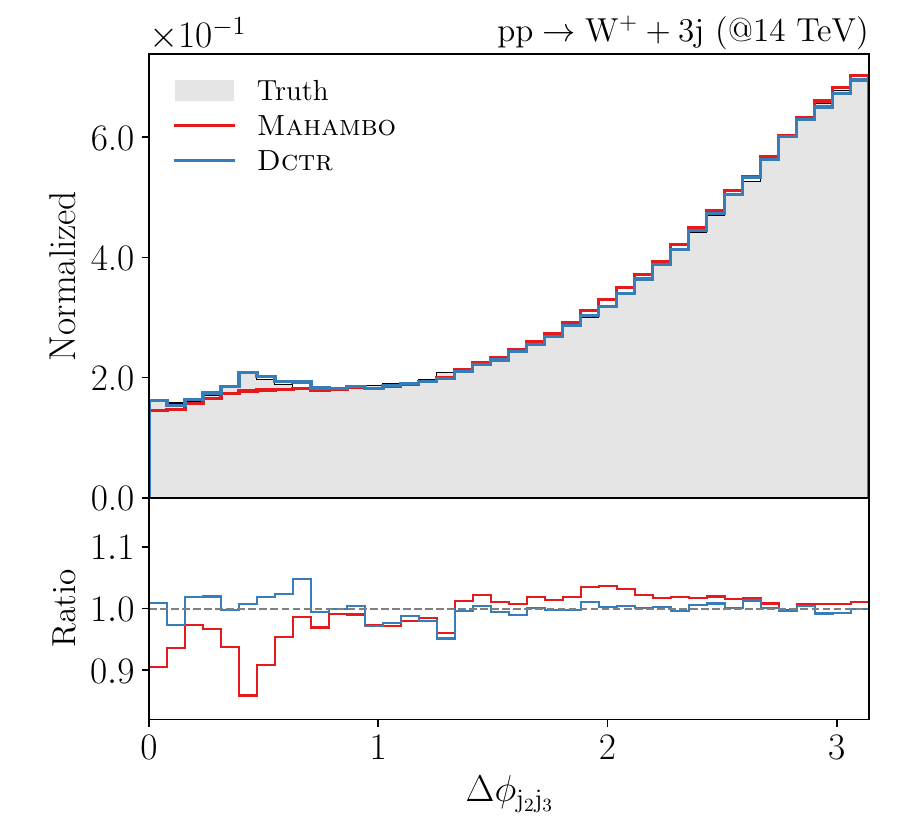}
    \includegraphics[page=16,width=0.32\textwidth]{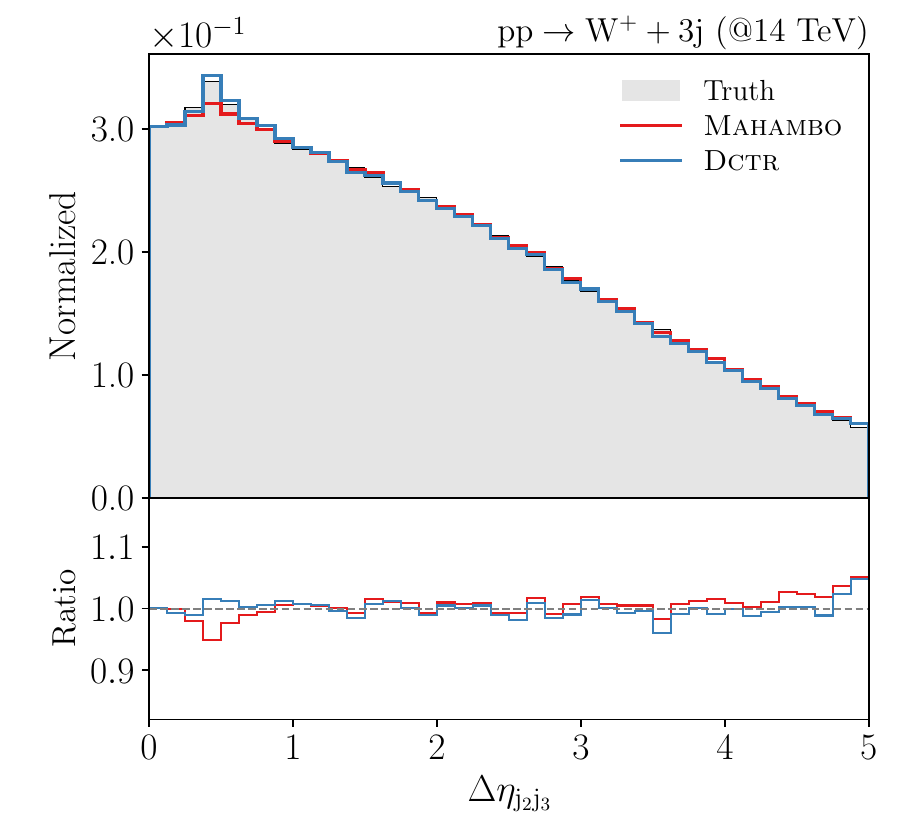}
    \includegraphics[page=17,width=0.32\textwidth]{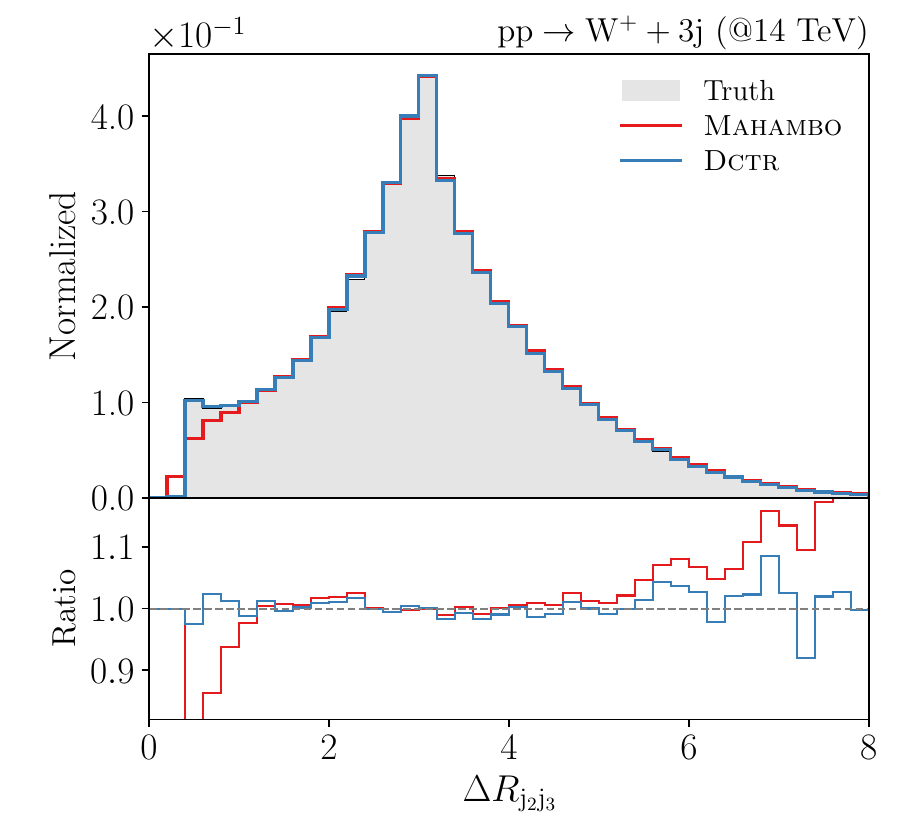}
    \caption{Transverse momentum distributions for the W-boson and the two leading jets (top)  and high-level angular diststibutions (bottom) for the $\Pp \Pp \to \PWp + 3\Pj$ process. We show the results for pure \mahambo  parametrization and including \dctr reweighting.}
    \label{fig:wp3j_mahambo_dct}
\end{figure*}

First, Fig.~\ref{fig:wp3j_mahambo_dct} demonstrates the reweighting in the data space when generating events in the \mahambo parametrization and feeding the data into the classifier in the \elfs parametrization. In this case, the weights defined by the classifier are no longer proper functions of the latent space, which naively does not allow to employ the \laser protocol. Since the data space and latent space are not having the same dimensionalities, we have to use the \omnifold protocol to pull back the weights to the latent space. However, it is as tricky for \omnifold to pull back the weights onto the latent space as obtaining useful classifier weights when training the classifier in the \mahambo representation. Apparently, it is difficult to detect the non-trivial deformations to the Gaussian latent space required to alleviate the small deficiencies in the \mahambo representation.

Instead, we can use these weights to reweight the data space directly, which is the bases of the \dctr protocol. As was observed in many other studies that employ classifier\-/based reweighting, including for ML\-/based simulation surrogate refinement~\cite{2009.03796}, we find that the weighted data\-/space is nearly indistinguishable from the truth and can capture sharp and subtle features.

\begin{figure*}[htbp!]
    \centering
    \includegraphics[page=3,width=0.32\textwidth]{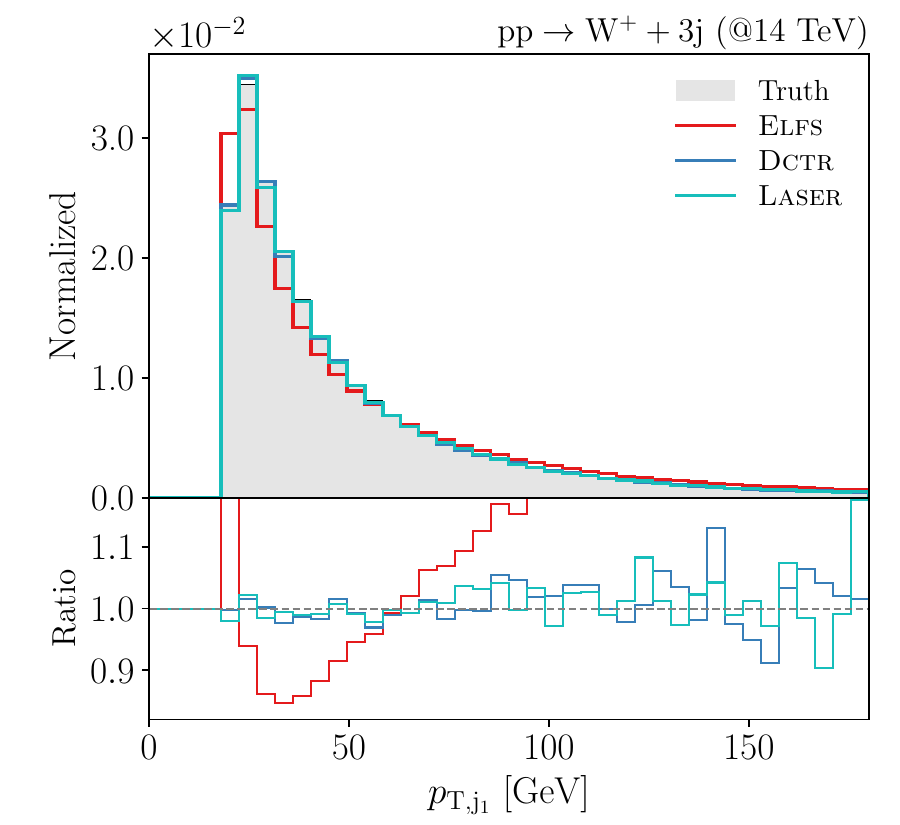}
    \includegraphics[page=9,width=0.32\textwidth]{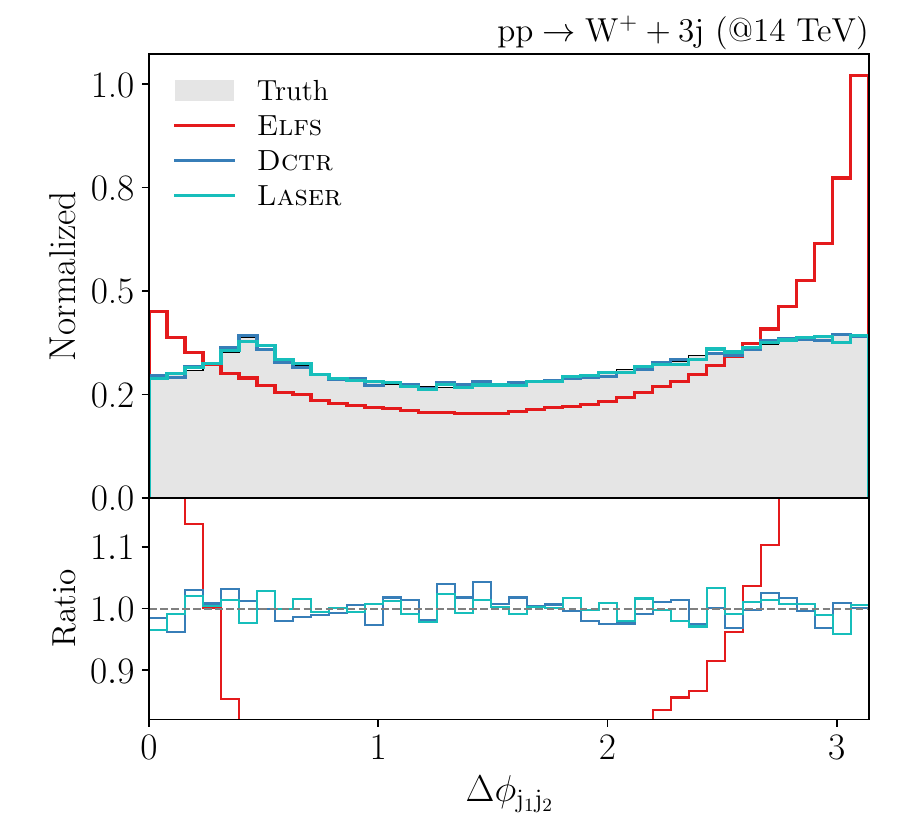}
    \includegraphics[page=11,width=0.32\textwidth]{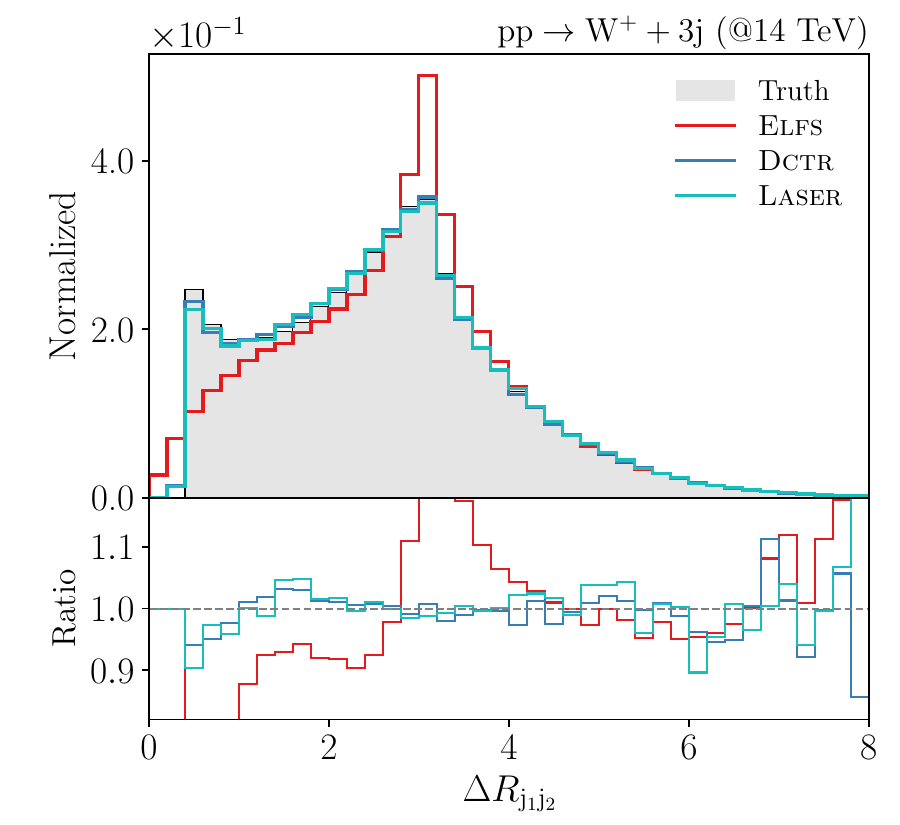}\\
    \includegraphics[page=15,width=0.32\textwidth]{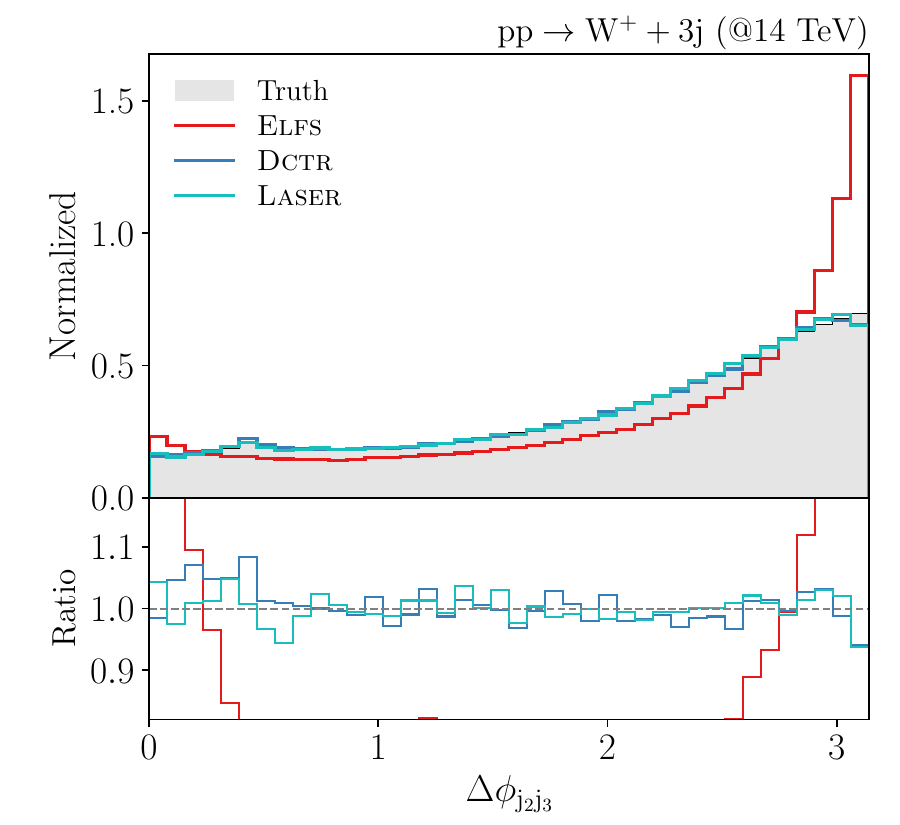}
    \includegraphics[page=16,width=0.32\textwidth]{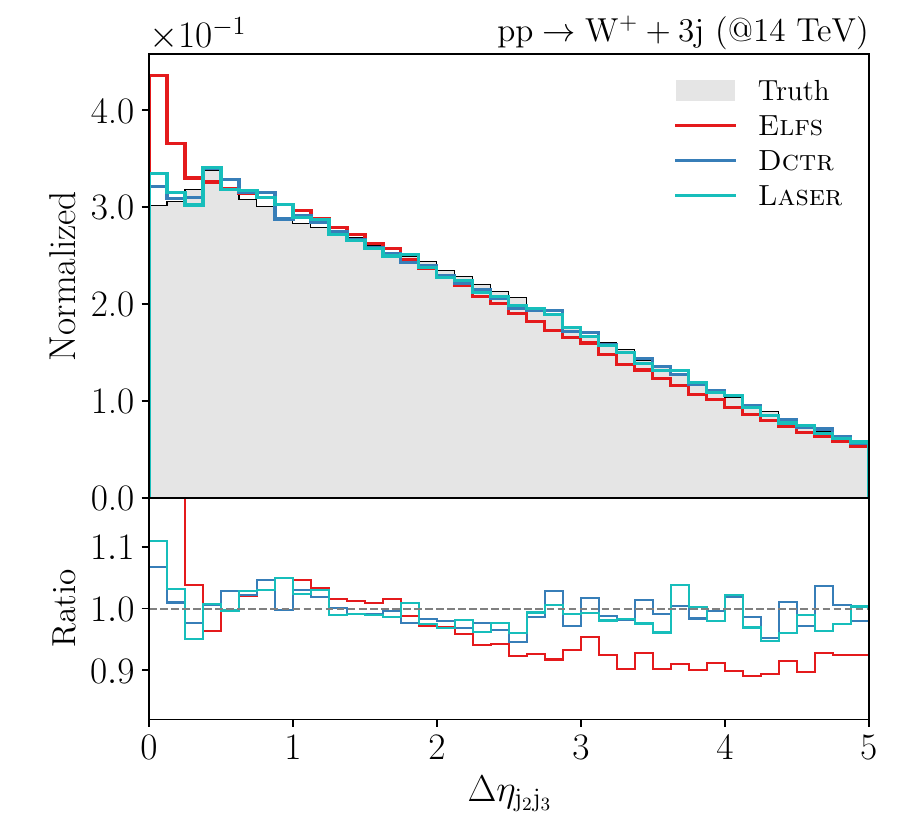}
    \includegraphics[page=17,width=0.32\textwidth]{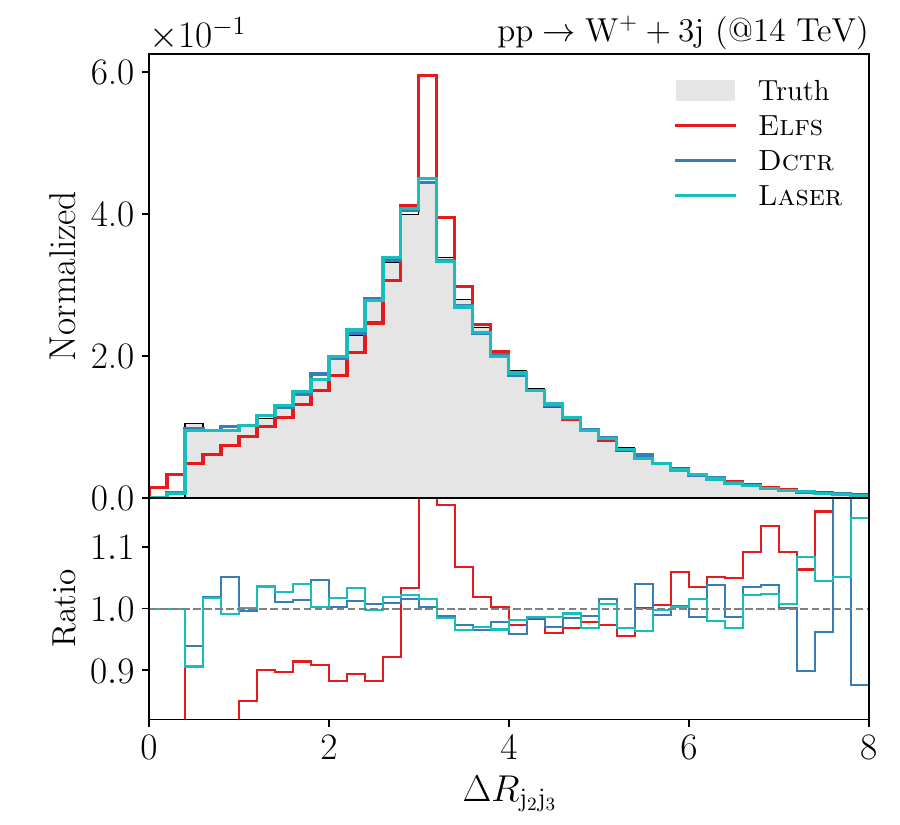}
    \caption{Transverse momentum distributions for the W-boson and the two leading jets (top)  and high-level angular diststibutions (bottom) for the $\Pp \Pp \to \PWp + 3\Pj$ process. We show the results for pure \elfs parametrization and including \dctr and \laser refinement.}
    \label{fig:wp3j_elfs_laser}
\end{figure*}

However, the weighted samples in the data space have less statistical power due to a variance in the weights. Figure~\ref{fig:wp3j_elfs_laser} explores how we can pull back the data-space weights to the latent space to produce unweighted examples in the data space. We start with \elfs instead of \mahambo as we found that pulling back \mahambo weights was ineffective as it requires \omnifold. Instead, if we use \elfs in both the generator and discriminator, the obtained weights are already proper weights of the latent space and allow using the \laser protocol. Without any refinement, \elfs cannot capture many of the key features of the phase space. However, \dctr works well, and the improvements directly translate almost verbatim when pulled back to the latent space and integrated into \laser. This includes all sharp and subtle features, especially in the variables that describe angles between jets. Even though \elfs starts further from the truth, the refinement makes it closer than \mahambo.

\section{Conclusions and Outlook}
\label{sec:conclusion}

Neural network surrogate models are entering an era of precision, where a qualitative description is replaced with quantitative agreement across the full phase space. Using $W$+jets matrix element simulations as an example, we explore how modifying the data before, during, and after generation can enhance the precision.  

Unsurprisingly, preprocessing is found to have a significant impact on downstream precision. Therefore, we proposed \mahambo, a variation of \rambodiet, to automatically reduce the data to its minimal\-/dimensional representation without needing process-specific interventions. This approach matched or outperformed other parameterizations across most of the phase space.

In agreement with previous 
studies, we find that classifier-based reweighting is able to precisely match the truth since the support of the probability density for our surrogate models is a superset of the truth.  

We also study two latent space refinement methods to avoid weighted samples. The best approach for the $\PW+\jets$ example was \laser, where the data-space weights are used in the latent space to produce unweighted examples.

While our primary focus has been on matrix element generation, the presented ML approaches have the potential for broader applications in the study of surrogate models. Furthermore, these techniques can be adapted for physics-based simulators, particularly in cases where refining the latent space allows for accessing the random numbers within the program. Finally, improving surrogate models to align accurately with physics-based simulations or observational data holds immense potential for enhancing various downstream inference tasks. We eagerly anticipate further exploration of these connections in future endeavors.

\section*{Code and Data}

The code and data for this paper can be found at
\url{https://github.com/ramonpeter/elsa}.
This allows the reader to either fully reproduce all our results with ease or adopt and implement our methods within other
software tools.

\section*{Acknowledgments}
\label{sec::acknowledgments}

We want to thank Tilman Plehn for fruitful discussions about classifiers and Theo Heimel for his help on the \prep parametrization. Further, we would like to thank Rob Verheyen for fruitful discussions about non-bijective flows and Marco Bellagente for his work earlier in this project. Finally, we thank Didrik Nielsen for his help on the \survae framework.
BPN was supported by the Department of Energy, Office of Science under contract number DE-AC02-05CH11231. RW was supported by FRS-FNRS (Belgian National Scientific Research Fund) IISN projects 4.4503.16. In addition, computational resources have been provided by the supercomputing facilities of the 
\foreignlanguage{french}{Université catholique de Louvain (CISM/UCL)} and the \foreignlanguage{french}{Consortium des Équipements de Calcul Intensif en Fédération Wallonie Bruxelles (CÉCI)} funded by the \foreignlanguage{french}{Fond de la Recherche Scientifique de Belgique (F.R.S.-FNRS)} under convention 2.5020.11 and by the Walloon Region.

\bibliographystyle{JHEP}
\bibliography{hep,HEPML}

\end{document}